\documentclass{aa}  
\usepackage{graphicx}
\usepackage{txfonts}
\usepackage[]{hyperref}
\usepackage{graphicx}	
\usepackage{amsmath}	
\usepackage{color}
\usepackage{soul}

\usepackage[T1]{fontenc}
\usepackage{ae,aecompl}

\usepackage{newtxtext,newtxmath}

\begin{document}

   \title{Random Forest classification of \emph{Gaia} DR3 white dwarf-main sequence spectra: a feasibility study}


   \author{David Echeverry
          \inst{1}
          \and
          Santiago Torres\inst{1,2}\thanks{Email;
    santiago.torres@upc.edu}
          \and
          Alberto Rebassa-Mansergas\inst{1,2}
          \and
           Aina Ferrer-Burjachs\inst{1}
          }

\institute{Departament de F\'\i sica, 
           Universitat Polit\`ecnica de Catalunya, 
           c/Esteve Terrades 5, 
           08860 Castelldefels, 
           Spain
           \and
           Institute for Space Studies of Catalonia, 
           c/Gran Capit\`a 2--4, 
           Edif. Nexus 104, 
           08034 Barcelona, 
           Spain}

\date{\today}

\titlerunning{Random Forest classification of \emph{Gaia} WDMS spectra}
\authorrunning{Echeverry et al.}

\offprints{S. Torres}

 
  \abstract
   {}
   {The third  {\it Gaia} data release  provides low-resolution spectra for around 200 million sources. It is expected that a sizeable fraction of them contain a white dwarf (WD), either isolated, or in a binary system with a main-sequence (MS) companion, i.e.  a white dwarf-main sequence (WDMS) binary. Taking advantage of a consolidated Random Forest algorithm used in the classification of WDs, we extend it to study the feasibility of classifying {\emph{Gaia}} WDMS binary spectra. }
   {The Random Forest algorithm is first trained with a set of  synthetic spectra generated by combining individual WD and  MS spectra for the full range of effective temperatures and surface gravities. Moreover, with the aid of a detailed population synthesis code, we simulate the {\it Gaia} spectra for the above mentioned populations. For evaluating the performance of the models, a set of metrics are applied to our classifications.}
   {Our results show that for resolving powers above $\sim$300 the accuracy of the classification depends exclusively on the SNR of the spectra, while  below that value the SNR should be increased as the resolving power is reduced to maintain a certain accuracy. The algorithm is then applied to the already classified SDSS WDMS catalogue, revealing that the automated classification exhibits an accuracy comparable (or even higher) to previous classification methods. Finally, we simulate the {\it Gaia} spectra, showing that our algorithm is able to correctly classify nearly 80\% the synthetic WDMS spectra.} 
   {Our algorithm represents a useful tool in the analysis and classification of real {\it Gaia} WDMS spectra. Even for those spectra dominated by the flux of the MS stars, the algorithm reaches a high degree of accuracy (60\%).}


   \maketitle
%

\section{Introduction}
\label{s:intro}

 The advent of large astronomical databases has surpassed the human capability of their direct analysis. Since the arrival of automated missions like {\it Hipparcos} in the 90's \citep{Turon1992}, large-scale projects such as the SuperCosmos Sky Survey \citep{Hambly1998}, the Sloan Digital Sky Survey (SDSS) \citep{York2000}, the Pan-STARRS collaboration
\citep{Kaiser2002}, the RAVE Survey \citep{Zwitter2008} or the Large Sky Area Multi-Object Fibre Spectroscopic Telescope (LAMOST) \citep{Zhao2012}, among other examples, have provided an unprecedented wealth and quantity of information. In particular, the current {\it Gaia} mission in its data release 3 (DR3; \citealt{Brown2021}), provides information for about nearly 2 billion sources. Data mining and machine learning methods and, in general, artificial intelligence techniques have therefore become an essential tool for handling such a large amount of data. In this sense, we can mention some pioneering works using these techniques in a wide variety of subjects, such as automatically discrimination of stars from galaxies \citep{Bazell1998}, classification of galaxies according to their morphology \citep{Naim1995}, estimation of the fraction of binaries in stars clusters  \citep{Serra-Ricart1996}, classification of main-sequence populations in the {\it Hipparcos} Input Catalogue \citep{Hernandez1994} or white dwarf (WD) classification in its Galactic components \citep{Torres1998}. Since then, numerous approaches have been proposed, mainly based on artificial neural networks, decision trees, discriminant analysis and many others \citep[see,][for a recent review]{Ball2010}. Among them, the Random Forest algorithm is one of the most promising techniques due to its versatility, robustness and easiness of implementation as we will demonstrate along this paper.

On the other hand, WDs are the most common remnant of intermediate stars ($\leq$ 8--11 M$_{\odot}$). Their theoretical properties are reasonably well understood \citep[e.g.][]{Althaus2010} and, as WDs can be very old objects, they can be used as reliable cosmochronometers carrying out a valuable information about the past history and evolution of our Galaxy \citep[see,][for a review]{GBerroOswalt2016}. However, two major drawbacks arise from the fact that WDs are high dense stellar objects. First, their large surface gravity leads to a broadening of the Balmer's spectral lines thus hindering accurate radial velocity determinations. Second, this same high surface gravity causes the sink of metals in the deep interior of WD atmospheres erasing any trace of the metallicity of the star. However, this lack of information can be retrieved if the WD is in a binary system, in particular, in a white dwarf plus a main-sequence (WDMS) star. In these cases, radial velocities  or metallicities can be indirectly derived from the WD companions \citep{Rebassa2016, Rebassa2021, Raddi2022}. Moreover, WDMS systems that arised from common envelope evolution are key to understand several important phenomena of our Galaxy, such as supernova type I progenitors, cataclysmic variables, low mass X-ray binaries and, in general, close binary evolution.

From an observational point of view, there exist several techniques for the identification of WDMS systems. If both objects are physically enough separated, a fact that happens in wide systems that usually avoided mass transfer episodes, they can be spatially resolved. Common-proper motion pairs identification is a widely use technique that can be applied in that situation. For instance, in the recent data provided by the {\it Gaia} mission in its early data release three (EDR3), around $16\,000$ WDMS systems have been identified in this way within 1\,kpc from the Sun \citep{ElBadry2021}. However, if the systems are close enough, indicating that they likely evolved through a common-envelope episode, they are hard to identify as long as the brightness of one of the components overwhelms the other.  In a recent study, \cite{Rebassa2021} identified 112 WDMS candidates in a nearly-complete volume sample up to 100\,pc from the Sun. The identification, based on the photometry of the objects and the accurate parallaxes provided by {\it Gaia} allowed to fit the Spectral Energy Distribution (SED) of both components and, thus, to derive its main stellar parameters. Although with this method is expected to retrieve 80\% of the unresolved WDMS binary underlying population, it is only circumscribed to a particular region of the Hertzsprung-Russell diagram. Outside this region, one of the components dominates over the other and it becomes impossible to individually identify the two components. The same authors estimate that this happens for $\sim$91\% of the systems. One way to overcome this problem is to make use of eclipsing systems, which reveal the presence of out-shined companions through eclipses in their light curves \citep{Pyrzas2012, Parsons2012, Parsons2017, Casewell2020}. Unfortunately, the percentage of eclipsing systems is low among WDMS systems \citep[$\simeq$10\,\%][]{Parsons2013}. Although spectra of unresolved WDMS are in principle also subject to these drawbacks, the use of artificial intelligence algorithms may help in differentiating single stars from binaries in which one of the components is out-shined.

In this sense, the largest spectroscopic WDMS binary catalogue to date has been obtained from SDSS and LAMOST data \citep[][and references therein]{Rebassa2016b,Ren2018}. It contains  $\approx4\,100$ systems ($\approx3\,200$ from SDSS and $\approx$900 from LAMOST). The identification, mainly based on a $\chi^2$-fit and a wave-let transform with a human visual inspection of individual spectra, results to be a robust and efficient method. However, it is expected  that {\it Gaia} will provide an equivalent number of WDMS spectra, thus implying a huge consumption of computational and human time. In this sense, automated classification methods become essential. Moreover, as mentioned above, WDMS binary spectra in which one of the components dominates the SED are nearly impossible to identify via $\chi^2$-fit/wavelet plus visual inspection methods, and it becomes important to evaluate whether or not artificial intelligence methods can help in overcoming this problem.

Machine learning techniques have been applied to spectral analysis \citep[e.g.][]{Bailer-Jones1998} and, in particular, Random Forest algorithms to the identification of stellar objects \citep{Perez-Ortiz2017,Plewa2018}. Regarding WDs, the Random Forest has been proved as an efficient tool in the classification of their Galactic components \citep{Torres2019} or their differentiation from MS stars \citep{Smart2021}.  Here we move a step forward and we aim to explore the capabilities of the Random Forest algorithm when applied to a continuous system, such as it is an spectrum, with a large (theoretically nearly infinite) number of features. Thus, in this paper we explore the possibilities of the Random Forest algorithm in the identification of WDMS systems through their spectra within the {\it Gaia} DR3 context. 

The paper is organized as follows. In Section \ref{s:rf} we introduce the Random Forest algorithm and define its main parameters. Section \ref{s:theo} is devoted to a general theoretical analysis of the Random Forest algorithm. The method is then  compared and checked in Section \ref{s:rfsdss} with a sample of already labelled objects provided by SDSS. In Section \ref{s:gaia} we study the capabilities of the method by emulating synthetic spectra for {\it Gaia} mission. Finally, the major conclusions and results are summarized in Section \ref{s:conclusions}.


\section{The Random Forest algorithm}
\label{s:rf}

The Random Forest algorithm is a well known supervised Ensemble Machine Learning method, widely used for classification purposes \citep{Breiman2001}. An initial labelled sample is needed in order to train the model. Once trained, the algorithm is tested in the labelled sample and if the accuracy of the model is good enough it can be applied to an unlabelled sample. The base of the Random Forest algorithm is the Decision Tree algorithm, in which the different branches are weighted with a
certain probability. The algorithm chooses the attribute from the
labelled data that best splits the subset by minimizing a certain function (typically an entropy or Gini index).  As a result of this training process, branches are weighted, thus being possible to apply the algorithm to an unlabeled sample for its classification. A detailed explanation of how the Random Forest works in the classification of the WD population can be found in \cite{Torres2019}. Additionally, for the reader who is not familiarized with Random Forest hyperparameters and definitions associated to the algorithm, we provide in Appendix \ref{a:hyper} a short description of the most relevant ones used in the present work.

\section{Theoretical classification of white dwarf-main sequence systems}
\label{s:theo}
\subsection{The synthetic sample}
\label{ss:synsamp}

In order to train the Random Forest algorithm we built synthetic samples of WDs, MS stars and WDMS binaries. We used the Phoenix MS star library \citep{Husser2013} and collected spectra of three different surface gravity values: 4.5, 5 and $5.5\,$dex. Their temperatures ranged from $2\,500$ to $4\,200\,$K in steps of $100\,$K, thus representing a total of 54 MS star model spectra.

Regarding WDs, we used the spectral library from \cite{Koester2010}, containing thirteen surface gravity values, from 6.5 to $9.5\,$dex in $0.25\,$dex steps. Their temperatures ranged from $6\,000$K to $10\,000$K in $250\,$K steps, from $10\,000$K to $30\,000$K in $1\,000$K steps, from $30\,000$K to $70\,000$K in $5\,000$K steps, and from $70\,000$K to $100\,000$K in $10\,000$K steps. In total, we ended up with 611 WD model spectra.

The best performance of the Random Forest is achieved for balanced data sets \citep{Breiman2001}. Consequently, we generated additional MS spectra by interpolating the temperatures for each surface gravity ranging from $2\,500\,$K to $4\,200\,$K in $6.25\,$K steps. In total we thus obtained 819 MS spectra, from which 615 uniformly distributed were  selected. 

For WDMS binaries there are not available simulated spectra, hence we created them by combining the fluxes from the MS and the WD model spectra previously introduced rescaled at a fixed distance of 200 pc. With the objective of having a balanced data set, only 729 were  generated with 27 MS and 27 WD stars uniformly distributed along all the range of temperatures and surface gravities. 

\begin{table}
\caption[]{Summary of the number of models and the range of parameters used to generate our synthetic spectra for the MS, WD and WDMS populations.}
\label{t:models}
         \begin{center}
     {\small
  \begin{tabular}{cccc}
  \hline
              \noalign{\smallskip}
 & MS population & WD population & WDMS population \\
            \noalign{\smallskip}
            \hline
            \noalign{\smallskip}

\# models & 615 & 611 &  729 \\
$T_{\rm eff}\,$(kK) & $2.5-4.2$ &$6.0-100$ & $2.5-4.2$ for MS \\
 " &  " & "  & $6.0-100$ for WD \\
 $\log g\,$(dex) & $4.5-5.5$ & $6.5-9.5$ &  $4.5-5.5$ for MS \\
 "  & "  & "  &  $6.5-9.5$ for WD \\
            \noalign{\smallskip}
            \hline
            \noalign{\smallskip}
  & & All populations & \\
            \noalign{\smallskip}
            \hline
            \noalign{\smallskip}
 $R$ & & $1-83\,000$ & \\
  SNR & & $0.25-100$ & \\
      \noalign{\smallskip}
 \hline
  \end{tabular}
}
\end{center}
\end{table}

\subsection{Pre-procesing and data preparation}
\label{ss:prepro}

The WD, MS and WDMS synthetic spectra created so far are virtually at an infinite resolution and are not affected by noise. Given that this is not the case for real spectra, we need to degrade our synthetic spectra according to a certain resolution and a given signal-to-noise ratio (SNR). 

When expressing the spectral resolution of an instrument, the concept of resolving power, $R$, is widely used, defined as:
\begin{equation}
    R=\frac{\lambda}{\Delta\lambda},
\end{equation}
where $\lambda$ is a considered wavelength, and $\Delta\lambda$ is the difference between peaks that can be distinguished, i.e. the resolution -- this is for example the measure used by the SDSS spectrographs \citep{Donald2000}, where the resolution $\Delta\lambda$ is calculated at full width at half maximum.

\begin{figure}
    \includegraphics[width=0.95\columnwidth,trim=0 0 20 0, clip]{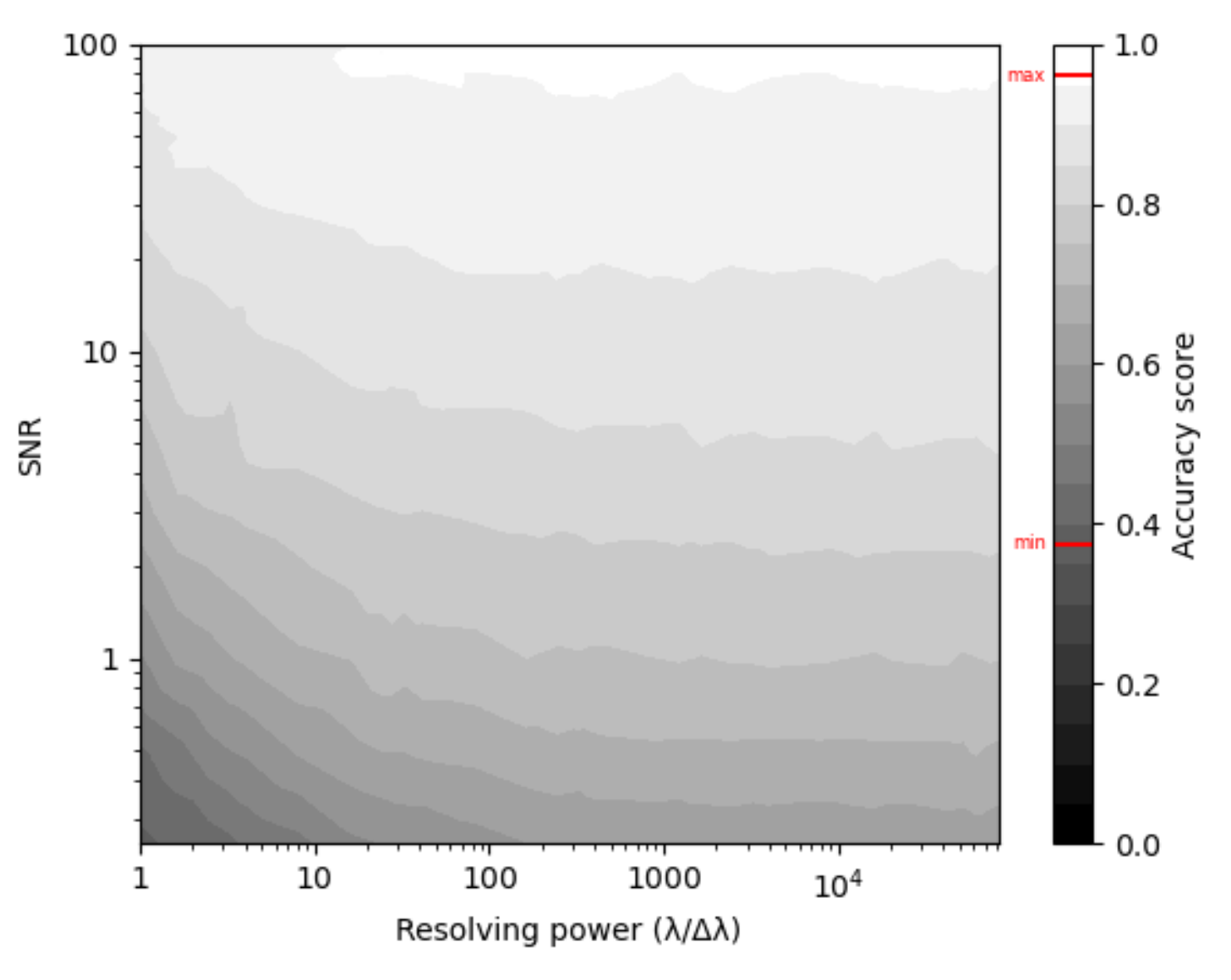}
                \caption{Accuracy density map as a function of $R$ and SNR. Minimum and maximum values of the accuracy are indicated in red in the gray scale.}
    \label{f:accuracy}
\end{figure}

The noise was introduced in the synthetic spectra by applying a white noise affecting all of the wavelengths and having a constant power spectral density (although a more specific noise model will be introduced for the SDSS case, see Section \ref{ss:sdss}, or the {\it Gaia} case, see Section \ref{ss:synspe} and Appendix \ref{a:noisemodel}). For this initial analysis, we adopted an additive white Gaussian noise (AWGN) centered at the signal value, $\mu_{\rm signal}$ and with standard deviation $\sigma_{\rm noise}$ which depends on the SNR. That is, the SNR for our synthetic model spectra is calculated as:
\begin{equation}
    {\rm SNR}=\frac{\mu_{\rm signal}}{\sigma_{\rm noise}}.
\end{equation}

Taking the previous definitions into account, we varied the resolving power from 1 to $83\,000$, and the SNR from 0.25 to 100. The resolution values are set to cover a wide range of resolving powers that could be possibly achieved by real spectrographs. The lower limits on the values of SNR and resolving power are chosen to analyse how the Random Forest performs with poorly resolved and noisy data. This analysis can therefore serve to future studies as an estimation of the possible classification results. A summary of  the models with the corresponding range of parameters analyzed in this work is shown in Table \ref{t:models}.

\begin{figure*}
    \includegraphics[width=0.67\columnwidth,trim=15 0 20 0, clip]{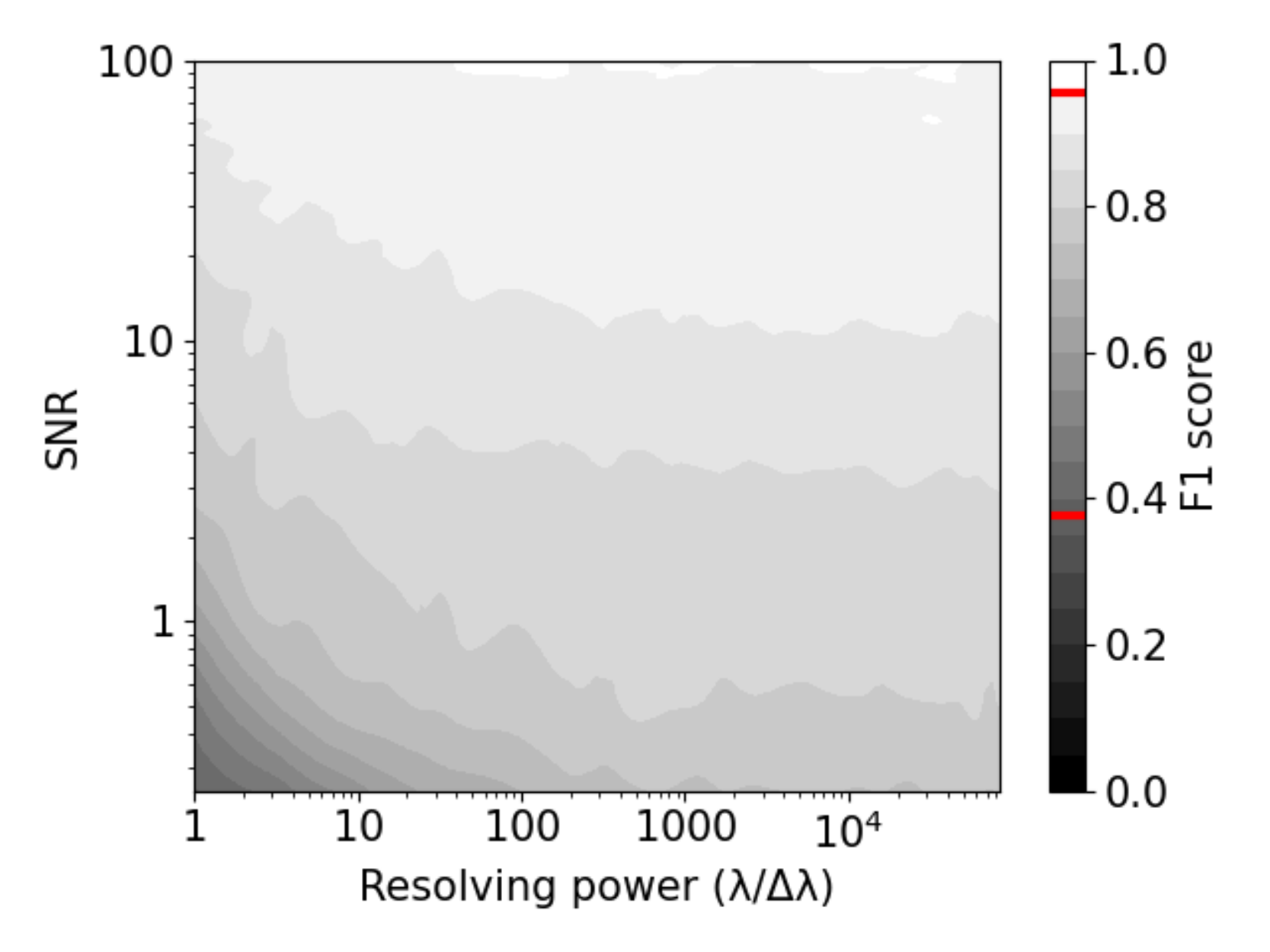}
        \includegraphics[width=0.67\columnwidth,trim=15 0 20 0, clip]{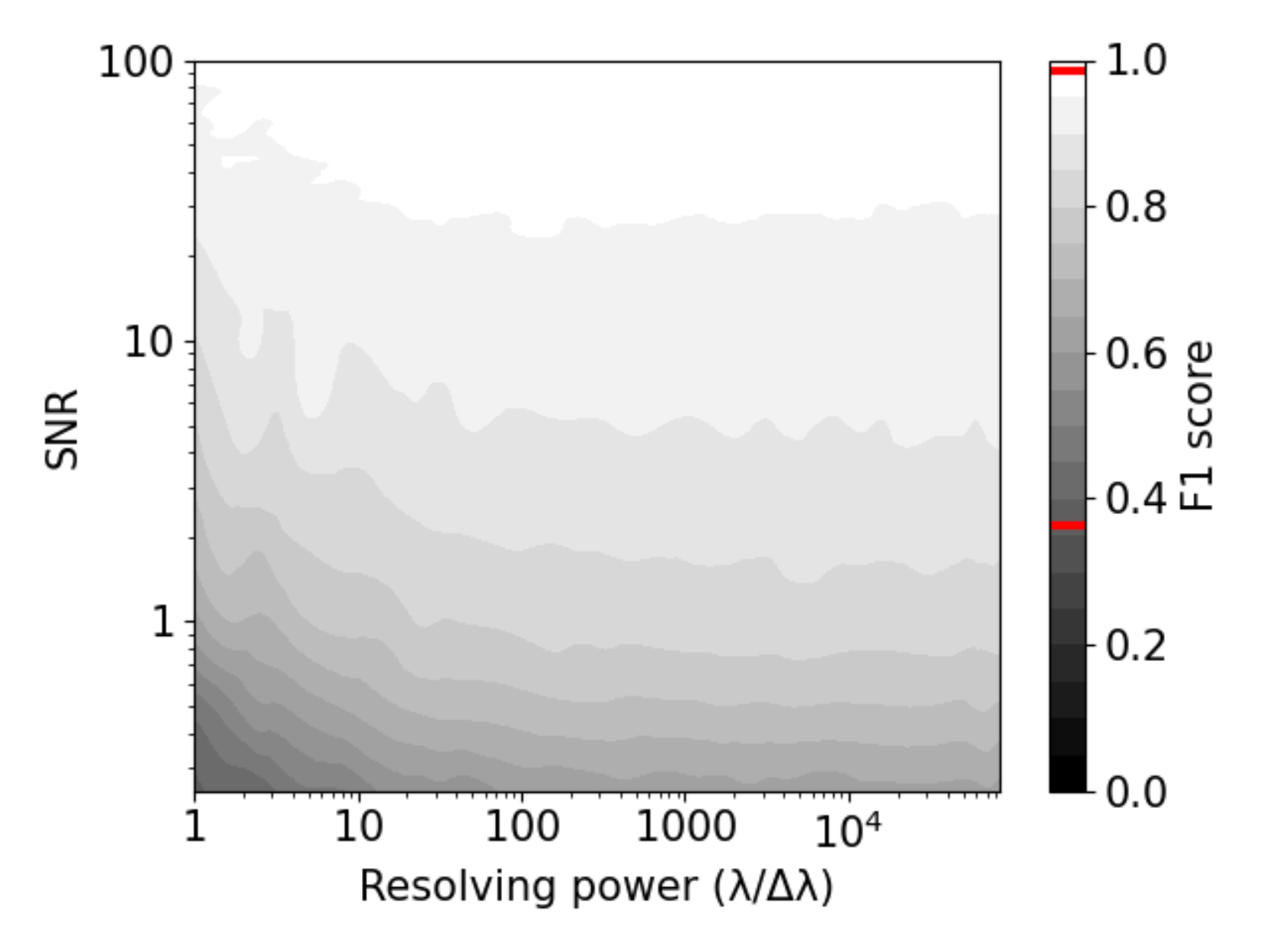}
                \includegraphics[width=0.67\columnwidth,trim=15 0 20 0, clip]{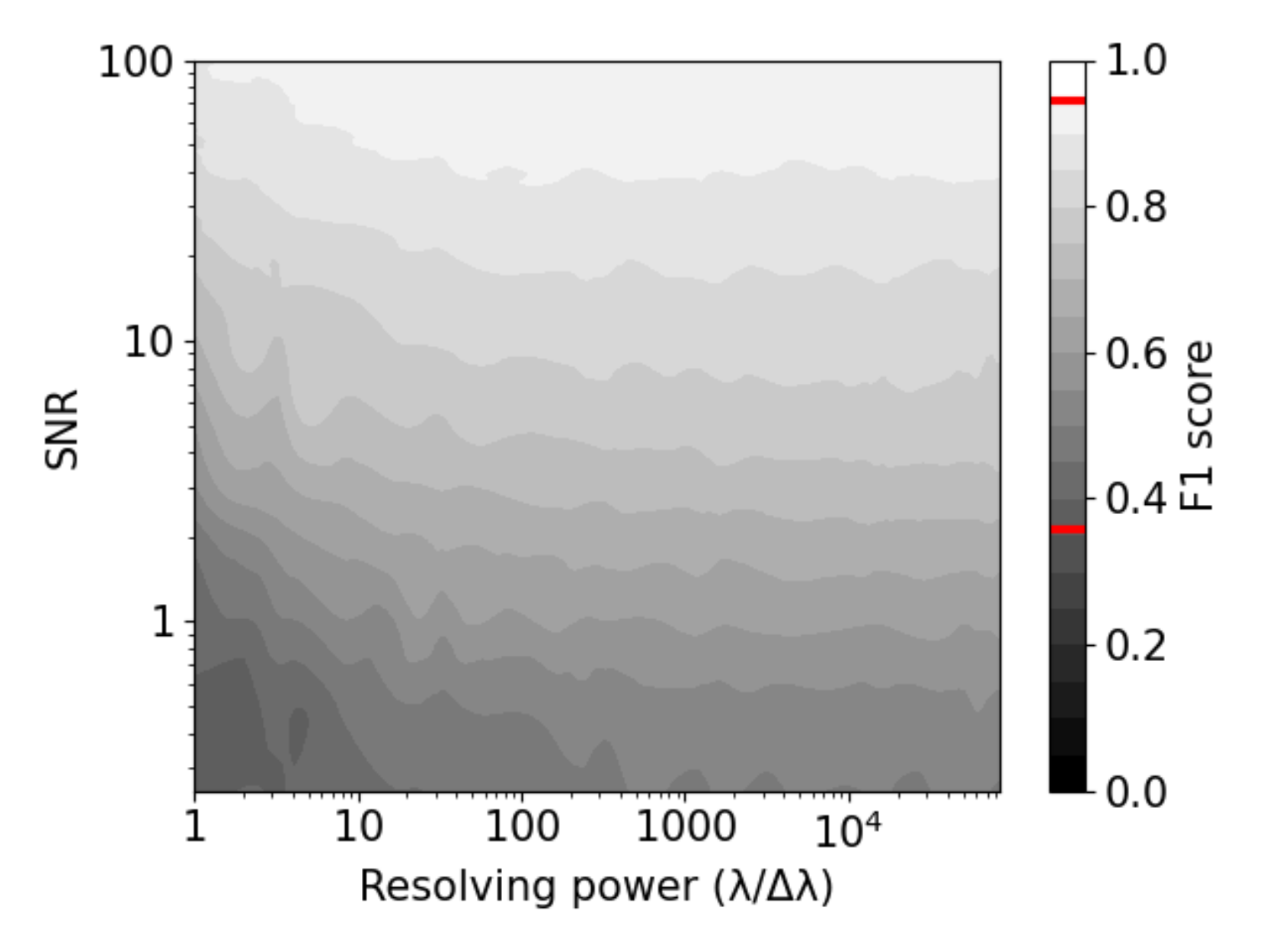}
   \caption{$F1$-score density maps as a function of $R$ and SNR for the MS (left panels), WD (central panel) and WDMS (right panel) populations. Minimum and maximum values of the respective score are marked in red in the gray scale.}
    \label{f:F1-maps}
\end{figure*}

\subsection{General performance as a function of $R$ and SNR}
\label{ss:general}

First of all we analysed the performance of the Random Forest algorithm when applied to the simulated synthetic spectra for different $R$ and SNR values. In particular, we varied $R$ from 1 to $83\,000$ for 30 evaluated points, and SNR in the range from 0.25 to 100 for 21 points. In total, we obtained 630 cases where the Random Forest performance was assessed, being the result of each one of those cases the average of 100 realizations.

The first metric  to be analysed is the general accuracy of the Random Forest algorithm. In Figure \ref{f:accuracy}, we show a density map of the accuracy (gray scale quantified in 5\% steps) as a function of $R$ and SNR.  

An initial inspection of Figure \ref{f:accuracy} reveals  that the accuracy ranges from a minimum value of 37.5\% for low $R$ and low SNR to a maximum of 96\% for high values of $R$ and SNR. In a closer look, we observe that for values of $R$ below $\approx$300, we need to increase the SNR as we decrease $R$ if we want to maintain a constant value of the accuracy. This implies that for low resolution spectrographs the higher SNR as possible is needed, otherwise the Random Forest algorithm will not be able to achieve an acceptable performance. On the contrary, for values of $R$ above 300 the accuracy score mainly depends on the SNR. This is a remarkable result, because if a Machine Learning algorithm is used for spectral classification purposes similar to those treated here in future scientific space missions, the use of a very high resolution spectrograph would not be required, since with a low resolution the same results seemed to be assured.

Complementary to the accuracy metric, in Figure \ref{f:F1-maps} we represent the F1-score (also known as balanced F-score) as function of $R$ and SNR for the MS, WD and WDMS populations (left, middle and right panels, respectively). This metric is the weighted average of the recall and the precision (see the Appendix \ref{a:hyper} for further details). In principle, we observe the same general trend in all three populations. That is, for $R$ values above $\approx 300$ the performance of the Random Forest is independent of the SNR. In particular, the WD population is the best classified by the algorithm.

\subsection{Performance at specific values of $R$ and SNR}
\label{ss:specific}

After analysing the overall performance of the Random Forest as a function of a wide range of $R$ and SNR values, we  focused on three specific cases:
\begin{itemize}
    \item Case 1, high resolving power and high SNR: $R=80\,000$ and SNR=100.
    \item Case 2, intermediate resolving power and medium SNR: $R=1\,800$ and SNR= 10.
    \item Case 3, low resolving power and low SNR: $R=40$ and SNR=1.
\end{itemize}

Case 1 represents the most ideal scenario in which we not only have high-resolution spectra but also high SNR. This resolving power can be achieved, for instance, at the Galileo Telescope located in La Palma using the HARPS spectrograph. The intermediate case (Case 2) is very similar to the spectra provided by the SDSS public data base (see Section \ref{s:rfsdss}). Case 3 covers the worse case scenario of both very low $R$ and SNR. This combination is unlikely in real observations but it is indicative of the lowest expectations the user can get from the Random Forest performance.

\begin{figure*}
    \includegraphics[width=0.68\columnwidth,trim=15 0 95 0, clip]{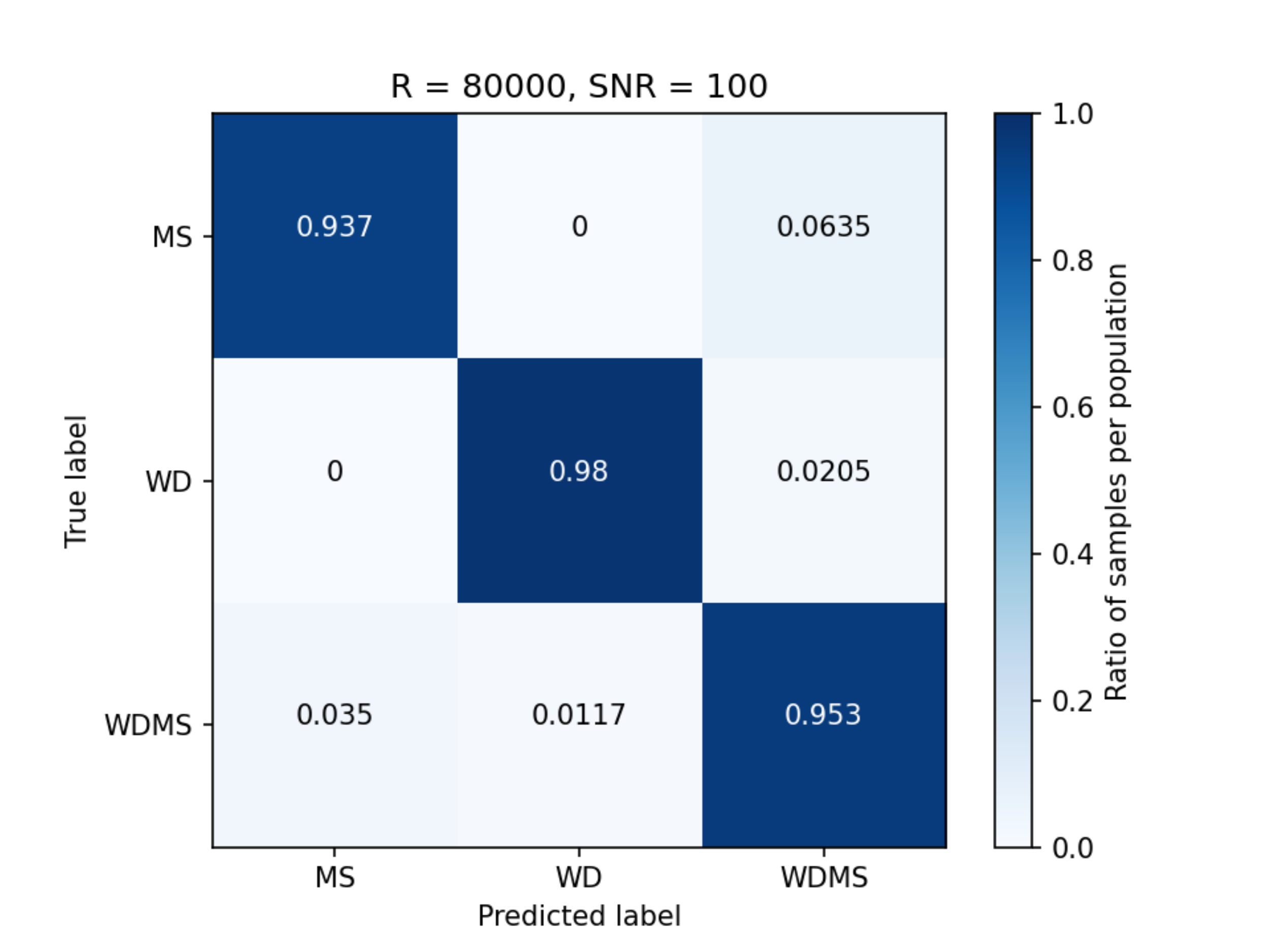}
        \includegraphics[width=0.68\columnwidth,trim=15 0 95 0, clip]{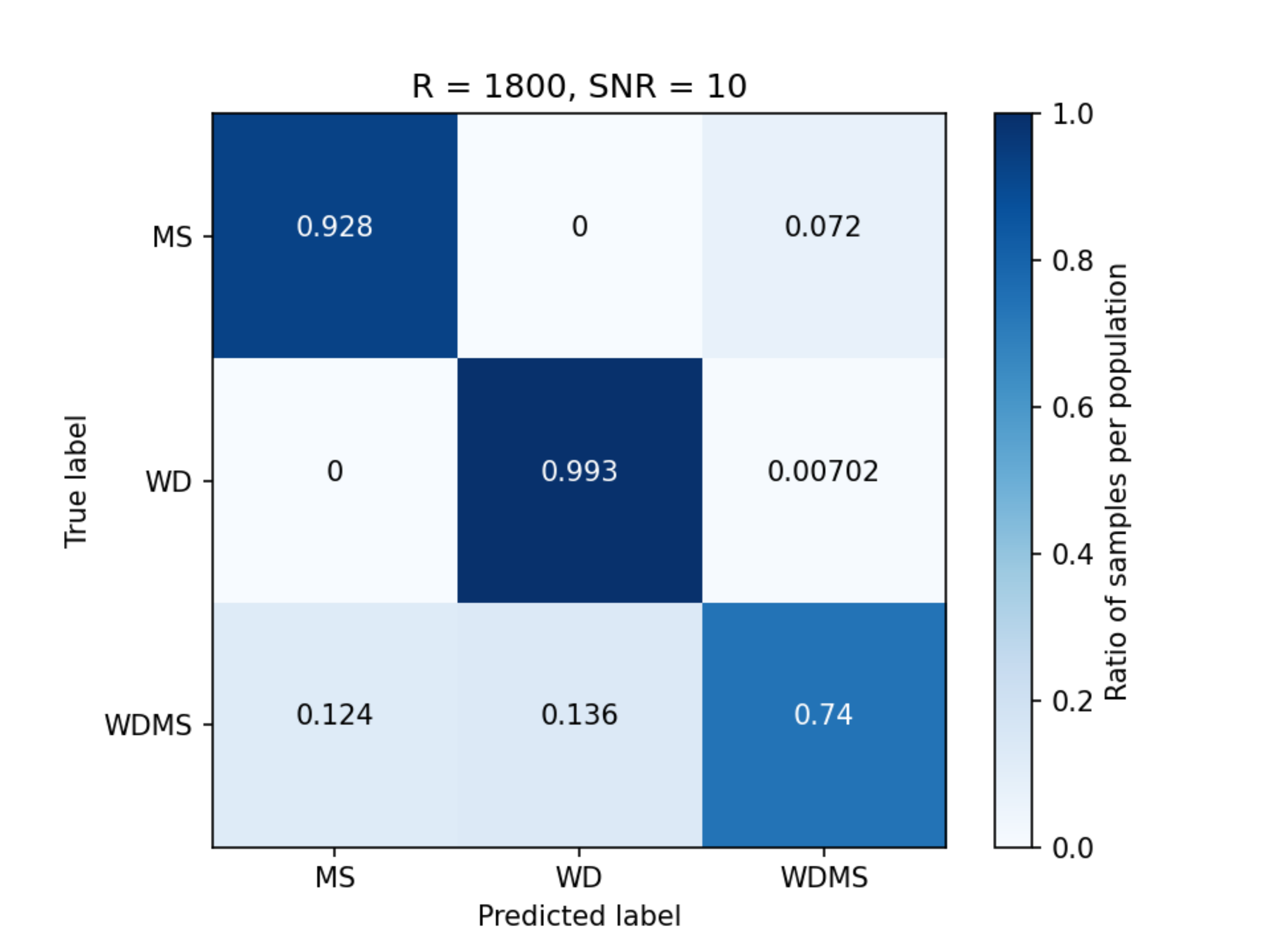}
                \includegraphics[width=0.68\columnwidth,trim=15 0 95 0, clip]{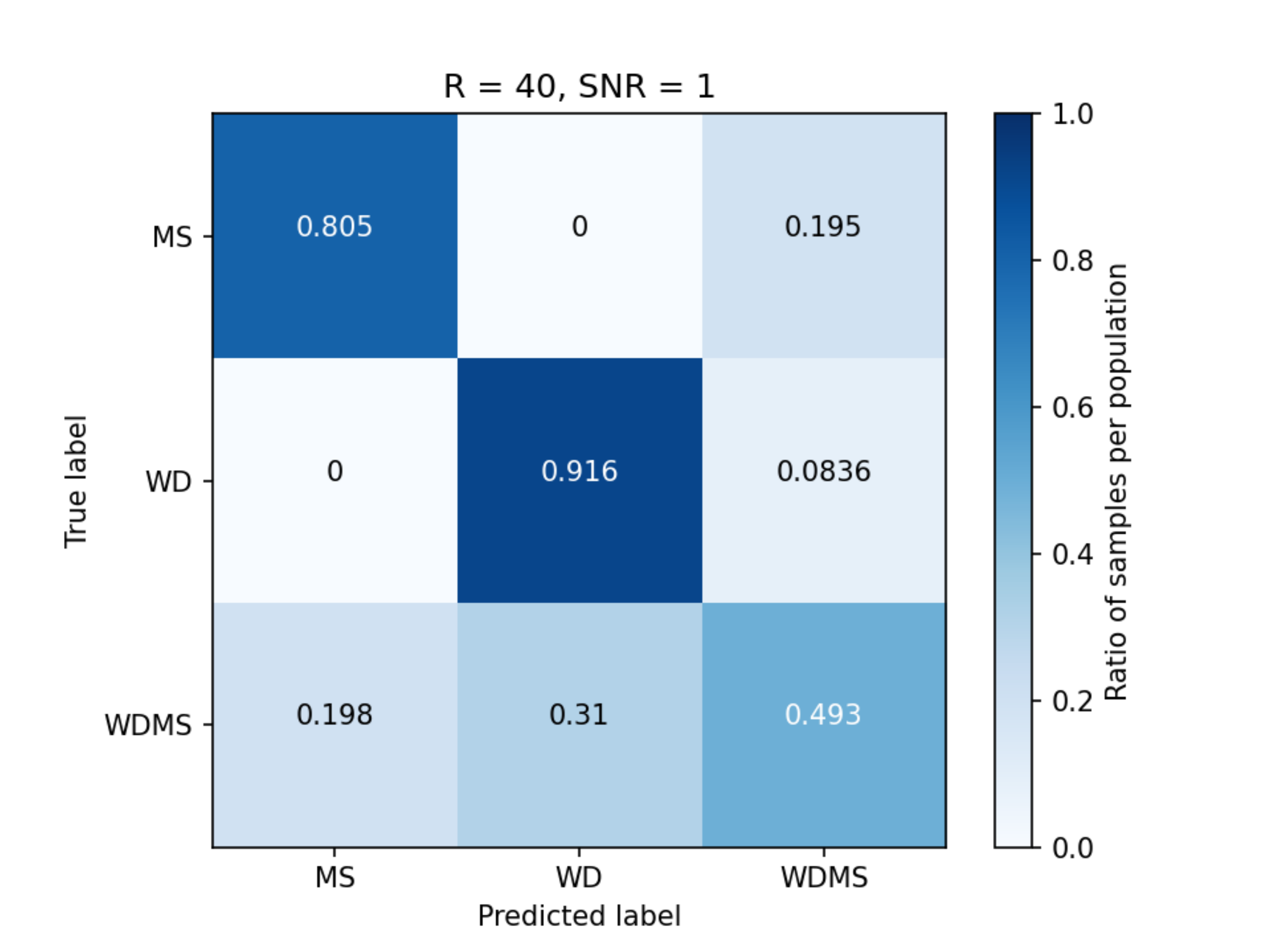}
                \includegraphics[width=0.68\columnwidth,trim=15 0 45 0, clip]{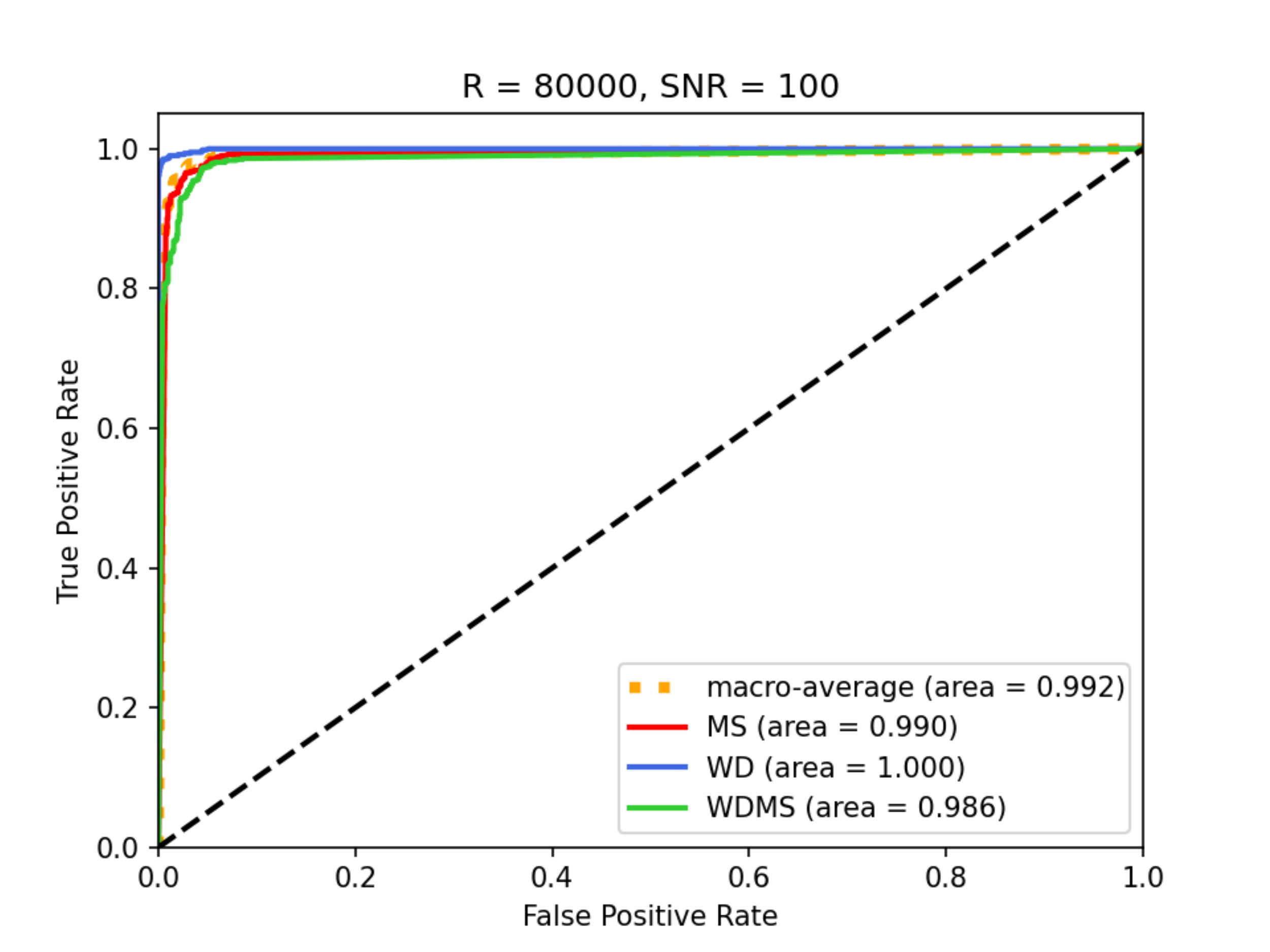}
        \includegraphics[width=0.68\columnwidth,trim=15 0 45 0, clip]{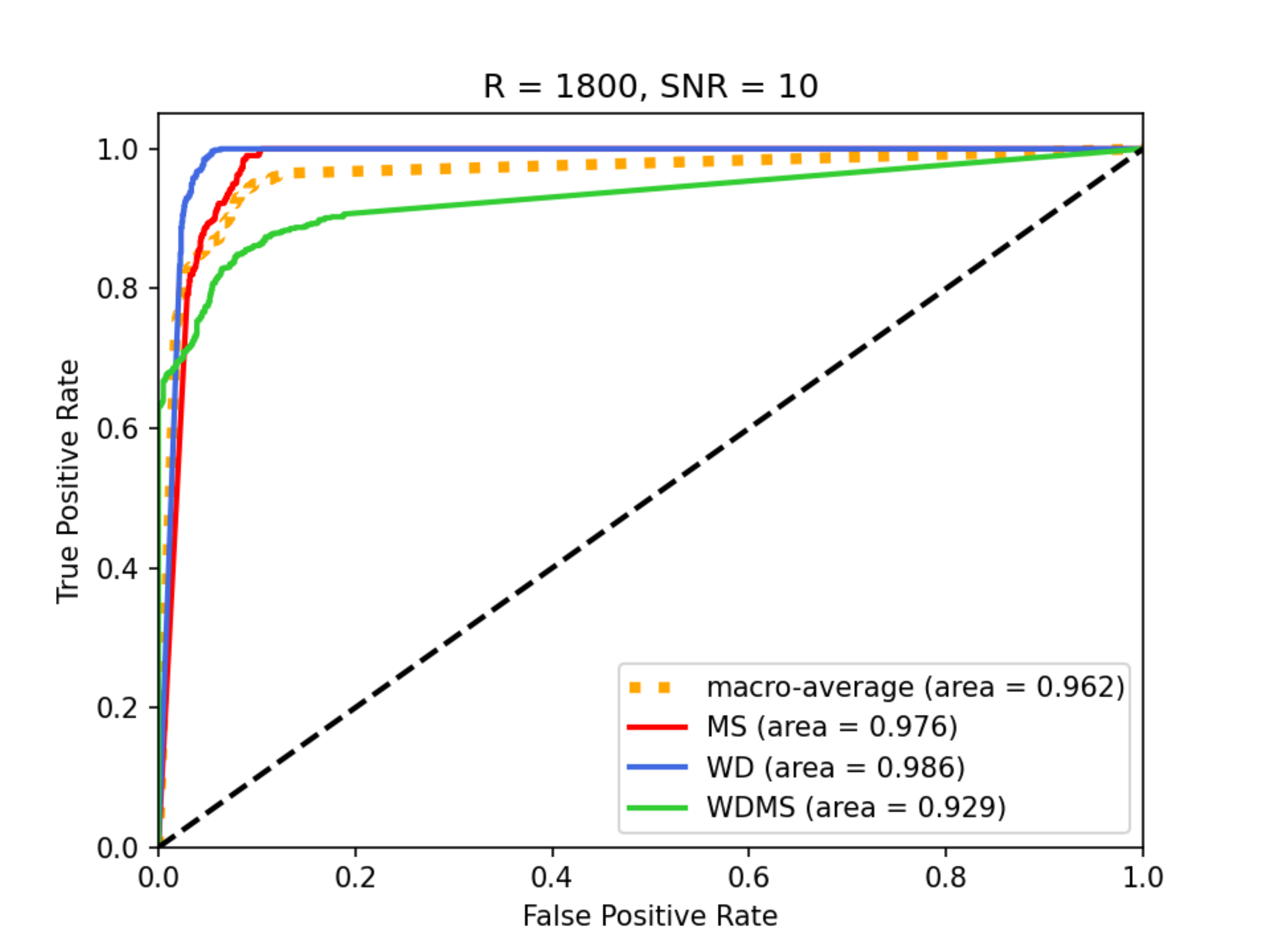}
                \includegraphics[width=0.68\columnwidth,trim=15 0 45 0, clip]{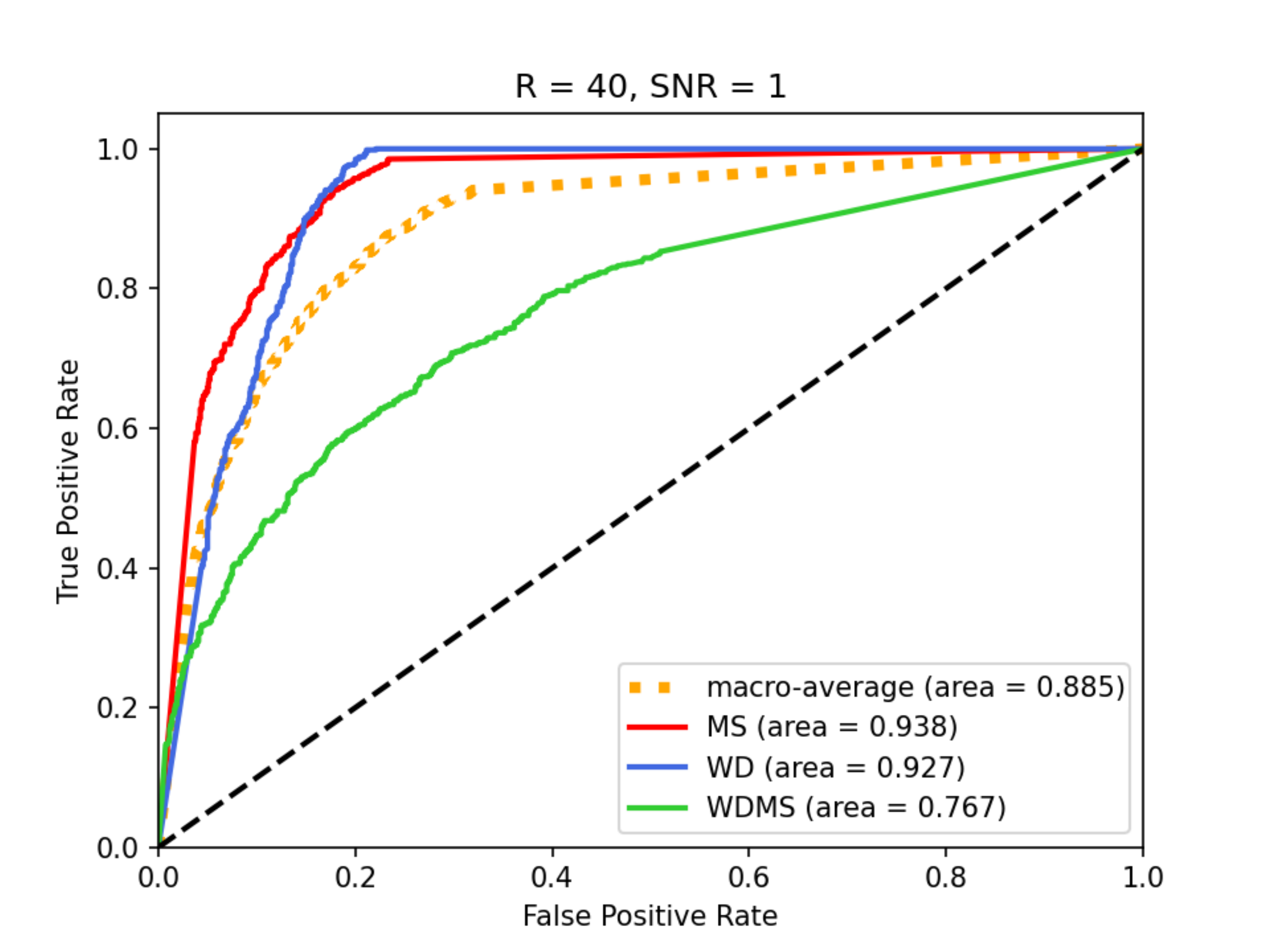}
         \caption{Confusion matrices (top panels) for the three cases under study and the corresponding ROC curves (bottom panels).}
    \label{f:cm3}
\end{figure*}

In Figure \ref{f:cm3} we show the confusion matrices (top panels) for the three cases under study and the corresponding ROC curves (bottom panels). The confusion matrices have been normalized by each population, that is, the numbers shown are the ratios of samples per population (the colour bar in the right provides this ratio). For the Case 1 (left panel),  high resolving power and high SNR, the algorithm performance is excellent. The population with the worst percentage of well-classified objects in this case is the MS population, with 93.7\% of the samples being correctly classified and only  6.35\% wrongly assigned to the WDMS population. Attending to WD stars, we have an almost perfect classification, with only 2\% of the population being mistaken as WDMS. For the WDMS population, the  ratio of well-classified objects is also very high, with only  5\% of misclassified samples: 3.5\% as MS and 1.17\% as WDs. This excellent performance is reinforce by the ROC curve (bottom-left panel), which we recall represents the True Positive and False Positive rates (TPR and FPR, respectively; see Appendix \ref{a:hyper} for further details). These rates are the recall or sensitivity of the algorithm and the probability of false alarm, respectively. We can check that for all three populations the curve is almost a 90-degree angle, which represents the perfect case.

Case 2 (middle panel), which represents a more common situation, continues to  provide an excellent classification for MS and WD stars, 93\% and 99.3\% correctly classified, respectively.  However, the percentage of well-classified WDMS decreases to 74\%. The missclassified spectra is almost equally distributed between the MS and WD populations, with around 12-13\% for each. This deterioration of the results is probably due to the lower SNR, according to the accuracy and F1-score maps of the previous section (see Fig. \ref{f:accuracy} and \ref{f:F1-maps}). Regarding the ROC curves for this case (bottom-middle panel), we can observe very sharp curves for the MS and WD populations as well as for the macro-average, but the AUC scores denounce substantial worse scores than in the previous case (see Appendix \ref{a:hyper} for a definition of the previous parameters). This is because even though the populations are very well classified, there is a certain rate of False Positives (WDMS classified as single stars). The WDMS curve is now visibly weaker than the others due to the worse performance of the Random Forest for this population.

Finally, we analysed the worst possible scenario, Case 3, with a low $R$ and SNR. Even in this case the confusion matrix (top-right panel) reveals for the single population a very high ratio of well-classified objects, with 80\% for the MS and a stunning 91.6\% for the WD population. This implies that even though the data are affected by a strong noise, the algorithm is still able to recover the single populations most of the times. It is not the same for the WDMS binaries, where the recall collapses to 50\%. However, this value is still larger than a random classification (showed in the ROC curves by a diagonal dashed line), which indicates that the algorithm is still able to extract some information from the spectra. It is worth noting that  30\% of miss-classified WDMS objects are labelled as WDs, while only 20\% are assigned to the MS category. This fact is indicative of how well the WDs are classified, prioritizing this population at the expense of the WDMS binary population, which is often mistaken.

\subsection{WDMS binary performance depending on the intrinsic stellar parameters}
\label{ss:intrinsic}

The Random Forest algorithm can provide us with a valuable bunch of information. For instance, we can recover the information of how WDMS binary systems are classified (or misclassified) according to the stellar parameters of their components, that is for example, the WD and MS effective temperatures and surface gravities.  

\begin{figure*}
\centering
    \includegraphics[width=1\columnwidth,trim=5 0 0 0, clip]{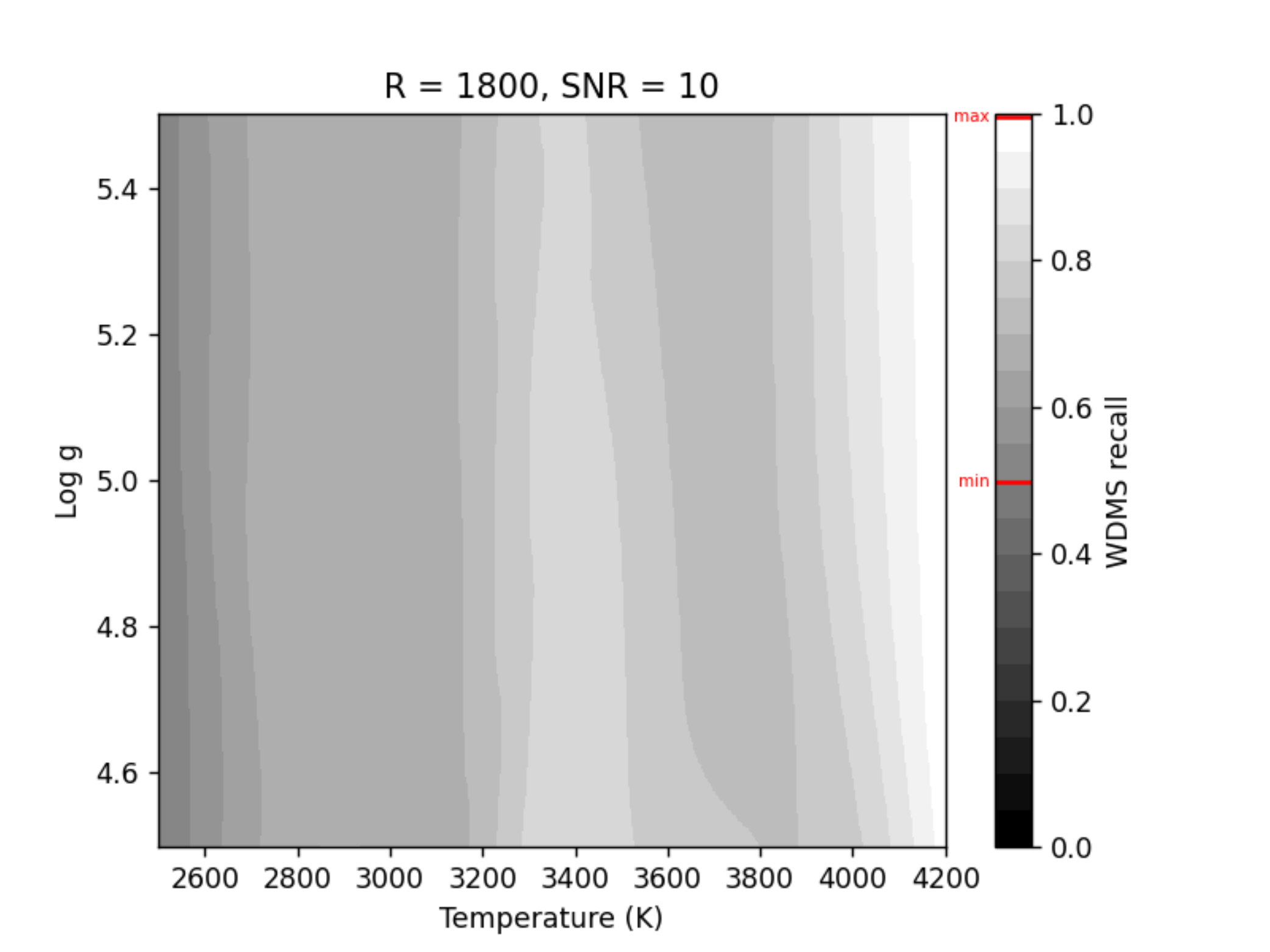}
        \includegraphics[width=1\columnwidth,trim=5 0 0 0, clip]{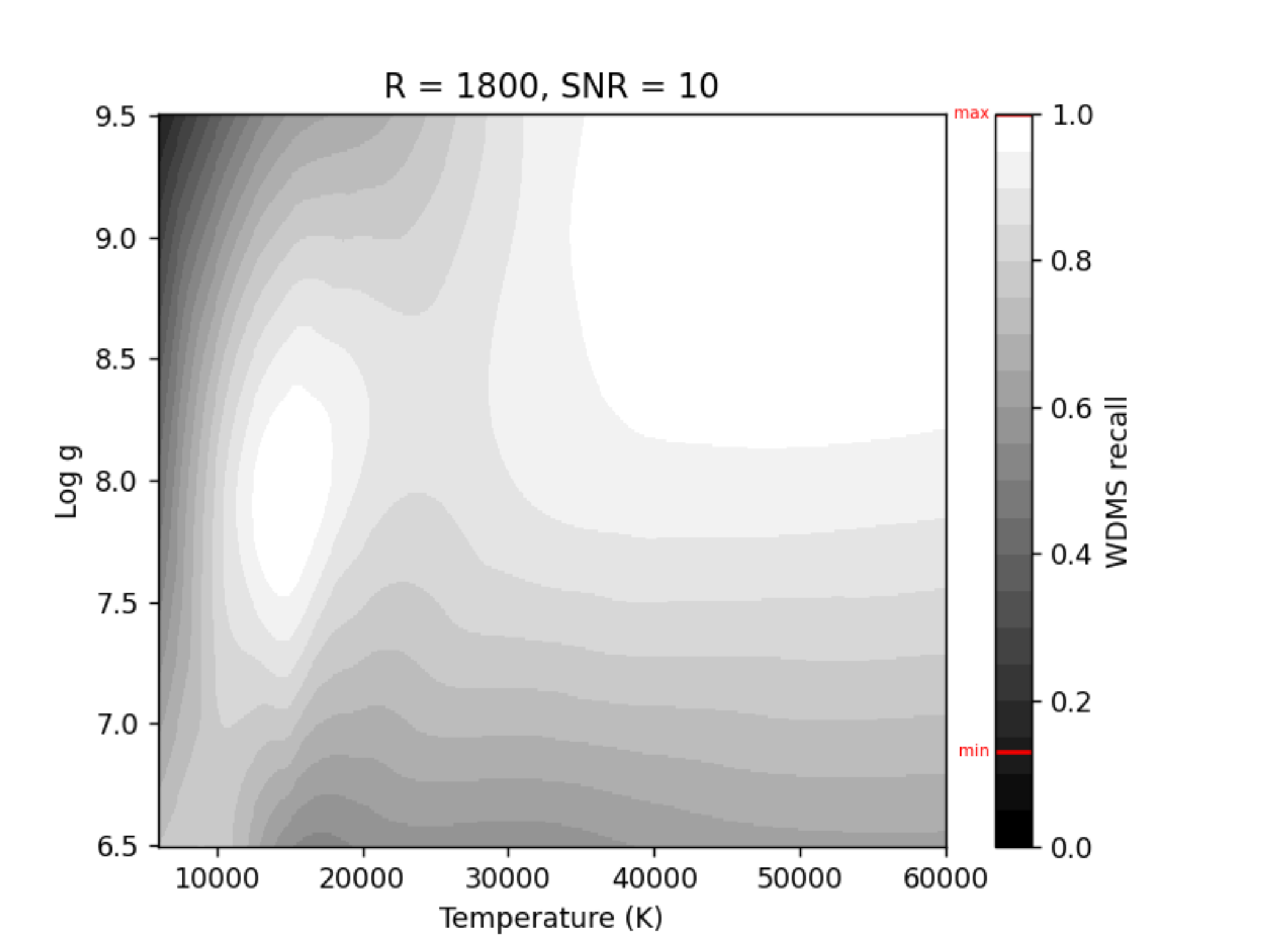}
    \includegraphics[width=1\columnwidth,trim=5 0 0 0, clip]{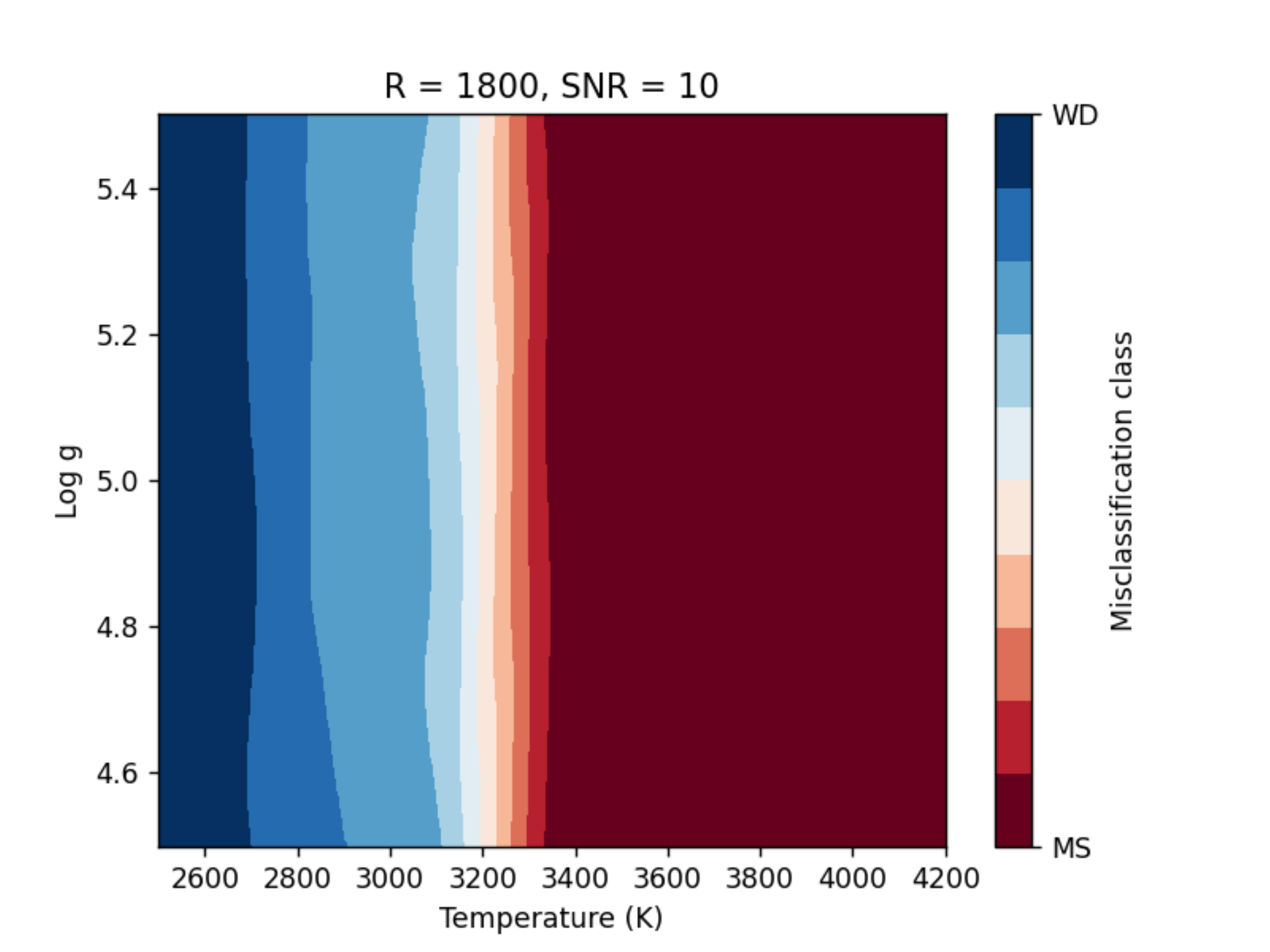}
        \includegraphics[width=1\columnwidth,trim=5 0 0 0, clip]{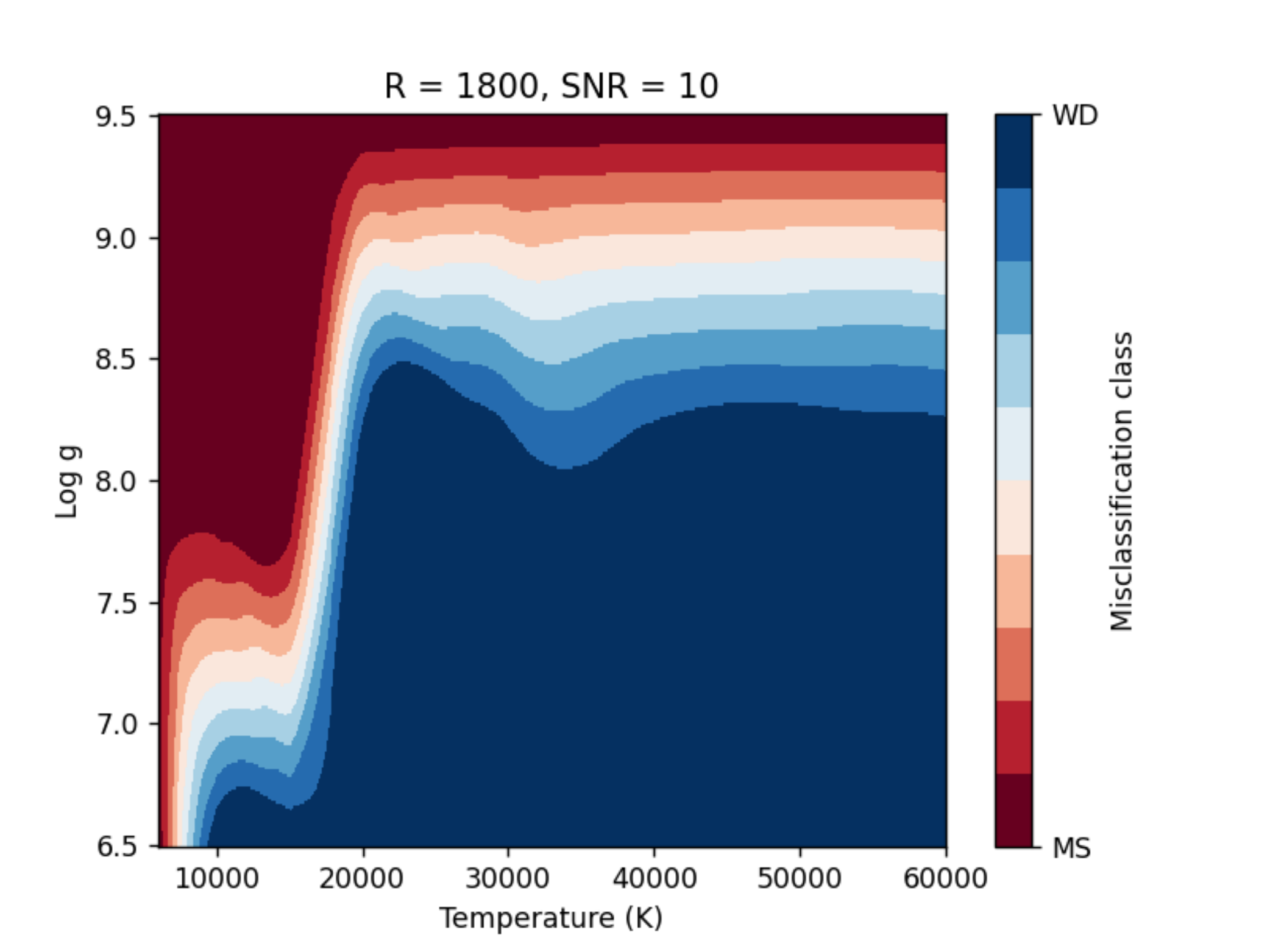}
                        \caption{Top panels: WDMS recall density map as a function of the MS (left panel) and WD component                      (right panel) stellar parameters. Bottom panels: WDMS miss-classification as a function of the MS (left panel) and WD (right panel) stellar parameters.}
    \label{f:intrin}
\end{figure*}

In the top panels of Figure \ref{f:intrin} we show the WDMS recall density map as a function of the MS component (left panel) and WD component (right panel) stellar parameters for our Case 2. The gray scale on the right of the panels indicates the WDMS recall (from 0 to 1 in steps of 5\%) and the minimum and maximum values are marked in red. In general, the recall score ranges between 50\% and 100\%  depending on the MS star parameters and between 10\% and 100\%  depending on the WD parameters. The maximum WDMS recall is achieved when the MS component reaches effective temperatures above $4\,100\,$K regardless of the surface gravity. This is because at these effective temperature values the main sequence components are K-type stars or earlier, which implies  the resulting WDMS binary spectra resemble the spectra of single MS stars. Since the training sample contains WDMS spectra dominated by the flux of the main sequence stars, all WDMS binaries with MS components hotter than 4\,100\,K ended up being correctly classified. On the other hand, WDMS are better recovered when the WD components have effective temperatures above $30\,000\,$K and surface gravities over 8.5 dex. This combination of parameter values,  result in WD spectra with fluxes that are of similar contribution to most of their MS companions, thus implying that the two components contribute significantly to the spectral energy distribution. The same situation takes place when considering WD components of effective temperatures between $\simeq$10\,000-20\,000\,K and surface gravities between $\simeq$7.5-8.5\,dex.

The darker areas, indicative of low recall, are concentrated at MS star effective temperatures below 2700\,K and WD effective temperatures below 10\,000\,K and surface gravities under 9 dex. In these cases one of the components is completely outshined by the other, leading to binary system spectra of very similar features as single MS stars or WDs. It is also interesting to see that for MS temperatures between $3\,200\,$K and $3\,500\,$K, the fluxes are on average comparable to the WD ones, so the resulting WDMS spectra are well classified. However, from $3\,600\,$K to $4\,000\,$K the fluxes of the main sequence spectra start to overtake the WD and, as a consequence, the algorithm miss-classifies the WDMS as MS stars. 

Complementary to the previous analysis, we show in the bottom panels of Fig. \ref{f:intrin} a density map indicating  to which wrongly class WDMS systems have been assigned. That is, in the bottom left panel we can confirm that WDMS are miss-classified as MS stars when their effective temperatures are above $3\,200\,$K, and as WDs for temperatures below that value regardless the surface gravity. On the bottom right panel, we observe in this case that the miss-classification depends as well on the WD surface gravity. When the WD component is cold, the WDMS systems are miss-classified as MS stars, however for effective temperatures above $10\,000\,$K and low surface gravity values, the binary system can be wrongly identified as a WD.

\subsection{Feature importance}
\label{ss:feature}

\begin{figure}
    \includegraphics[width=0.95\columnwidth,trim=-10 0 20 0, clip]{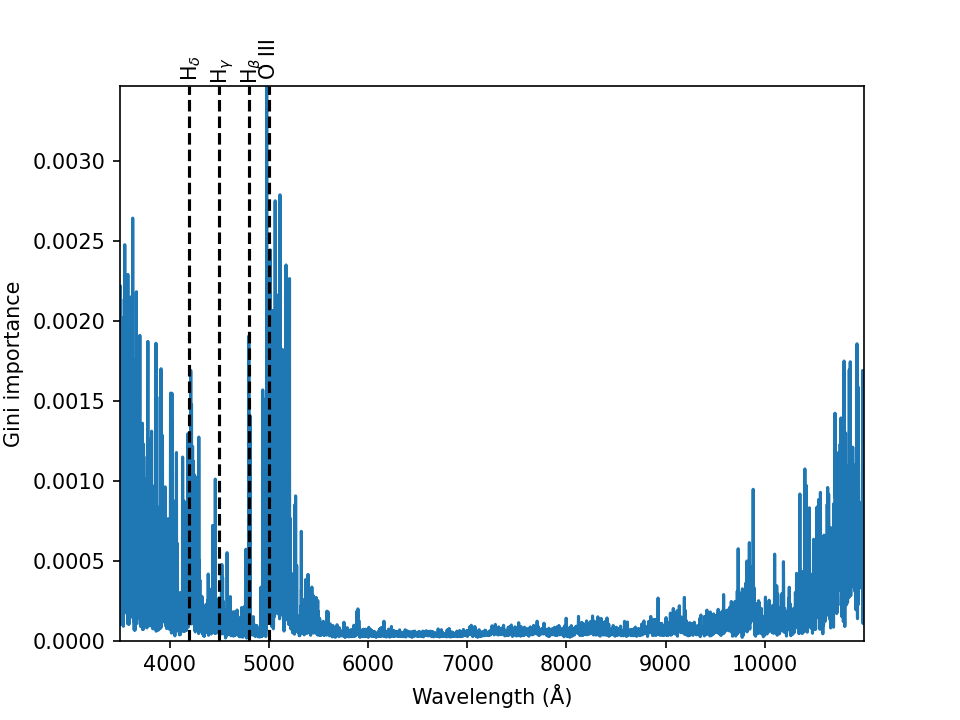}
                \caption{Feature importance as function of the wavelenght for our Case 2. As a visual reference we marked as vertical dashed lines the doubly ionized oxygen transition O III, and the H$\delta$, H$\gamma$ and H$\beta$  Balmer lines transitions on WDs.}
    \label{f:importance}
\end{figure}

The Random Forest algorithm through its ensemble of decision trees assigns a Gini function (or also equivalently an entropy function; see Appendix \ref{a:hyper}) to each feature (in our case, wavelength). Thus, after the training process has been accomplished, we can deduce which of these features  have been mastered the classification process, that is, the feature {\sl importance}. In Figure \ref{f:importance} we plot the  Gini function importance as a function of  wavelength for our Case 2.  The first thing to remark is the higher importance of the extreme wavelengths, corresponding to the blue and red limits of the spectra. This is a consequence that both types of stars, WD and MS, generally emit more at red and blue wavelengths, respectively. However, the larger importance value is achieved for wavelengths around $5\,000\,$\AA. This wavelength can be related to a common transition found on MS stars: the doubly ionized oxygen transition or O III. Additionally, some other relevant wavelengths have been found in the range from $4\,000$ to $5\,000\,$\AA. We clearly identify the H$\delta$, H$\gamma$ and H$\beta$ Balmer lines, located at $\simeq$4\,100, $\simeq$4\,300 and $\simeq$4\,900\,\AA,  respectively.

Finally, and additional implication can be derived from Fig. \ref{f:importance}. As observed, there exist three ranges of wavelengths located at $3\,500$, $5\,000$ and $11\,000\,$\AA, which account for the maximum information in the classification process. This would imply that three photometric pass-bands covering these wavelengths -- for instance, Sloan (u’,\,g’,\,z’), Johnson-Cousin (U,\,B,\,I) or Hubble Space Telescope (F336W,\,F450W,\,F814W) --, may also provide a good solution in the classification of WD, MS and WDMS populations.

\section{Random Forest classification of SDSS spectra}
\label{s:rfsdss}

So far we have analysed the performance of the Random Forest for classifying synthetic spectra. The next step is to test the algorithm with observed, already labelled, data. We take advantage of the observed data collected by the SDSS belonging to MS, WD and WDMS populations.

\subsection{The SDSS spectra}
\label{ss:sdss}

The Sloan Digital Sky Survey (SDSS) \citep{York2000, Gunn2006, Wilson2019} is one of the most detailed and largest astronomical surveys ever made, reaching more than three million of observed spectra. SDSS spectra are obtained with a 2.5\,m optical telescope located at the Apache Point Observatory in New Mexico using a pair of fiber-fed double spectrograph covering the wavelengths from $\sim$3\,800\,\AA\; to $\sim$9\,200\,\AA\; with a resolving power of 1\,800$\sim$2200. The SDSS contains the most complete spectroscopic catalogues of MS, WD and WDMS binaries \citep{West2011, Kepler2021}, which were classified in the case of WDMS binaries thanks to different techniques that involve human supervision \citep{Rebassa2007, Rebassa2010, Rebassa2012, Rebassa2016b}.

\begin{figure}
    \includegraphics[width=0.95\columnwidth,trim=-10 0 20 0, clip]{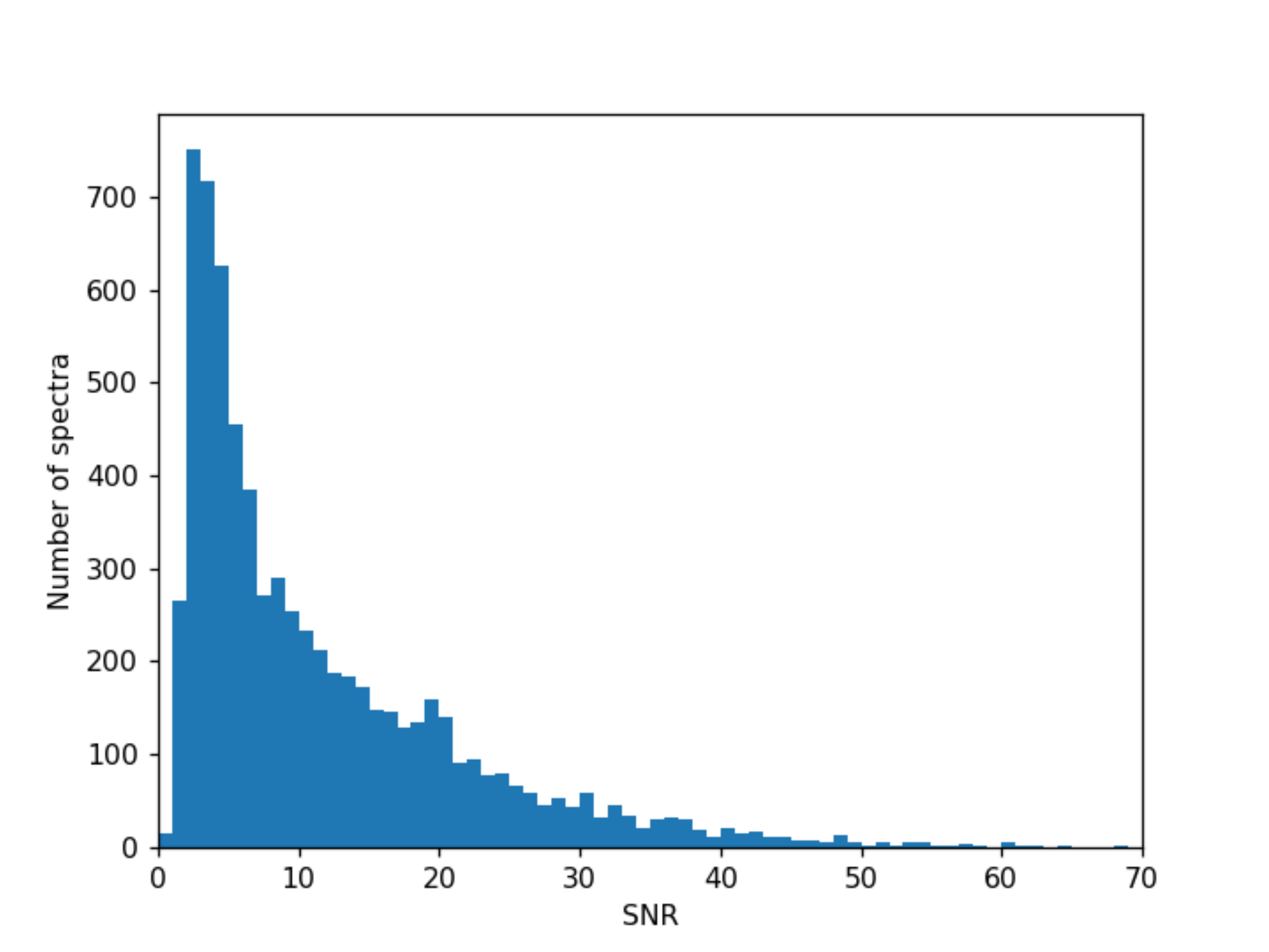}
                \caption{SNR histogram distribution of the SDSS spectra.}
    \label{f:SDSShisto}
\end{figure}

From the available catalogues, we extracted a practically balanced sample containing $2\,340$ MS, $2\,031$ WD and $2\,552$ WDMS binary spectra --- that is, a total of  $6\,923$ labelled spectra. These samples were selected in a way that contained spectra representative of all possible stellar parameter values. The resolving power is around $1\,800\sim2\,200$ depending on the used spectrograph, while the wavelength range of the spectra can vary slightly among the spectra. In order to normalize the sample we introduced a cut from $3\,850$ to $9\,150\,$\AA. Hence, we obtain a reliable sample where to validate the performance of our Random Forest algorithm. All SDSS spectra have associated flux errors, which can be used to derive an approximate value of the SNR for each spectrum as follows:

\begin{equation}
    {\rm SNR} = \frac{\sum\limits_{i}^{n} \frac{flux_i}{error_i}}{n},
\end{equation}

\noindent where $i$ refers to each one of the wavelengths available in the spectrum, $flux_i$ is the value of the flux for that wavelength and $error_i$ the error estimation at that certain wavelength. The total number of wavelengths is represented by $n$, where $n=3\,760$.  The range of SNR for the chosen SDSS spectra goes from 0.45 in the worst case to 100. In Figure\,\ref{f:SDSShisto}  we depict the SNR distribution for our selected SDSS objects. Most of them have SNRs below 50, with an average at around 11. It is worth noting that there is a large fraction of objects with a SNR as low as 5. This fact could be a problem when classifying the spectra, as we have seen in Section \ref{ss:general} that with a resolving power higher than 300 the performance of the algorithm only depends on the SNR.

\subsection{Classification results}
\label{ss:clas}

All the previously observed selected objects constitute the testing set of the classification process. That is, they are already labelled objects that we want to reclassify using our Random Forest algorithm. We adopted a training set formed by simulated spectra at a resolving power of 1800, typical of SDSS, and a set of 10 different values of SNR, 1, 3, 5, 7, 10, 15, 20, 40, 60 and 100,  in coherence with the observed spectra. Therefore, from the starting 615 MS, 611 WD and 729 WDMS synthetic spectra -- $1\,955$ in total --, we obtained a training set composed of $19\,550$ simulated spectra. By doing this, we assure that the algorithm is fed with all of the possible spectra that are expected to be found in the SDSS observed sample. In each case we added white Gaussian noise as explained in Section\,\ref{ss:prepro}, with a null mean and a standard deviation according to the SNR. 

\begin{figure*}
\centering
    \includegraphics[width=0.9\columnwidth,trim=0 0 10 0, clip]{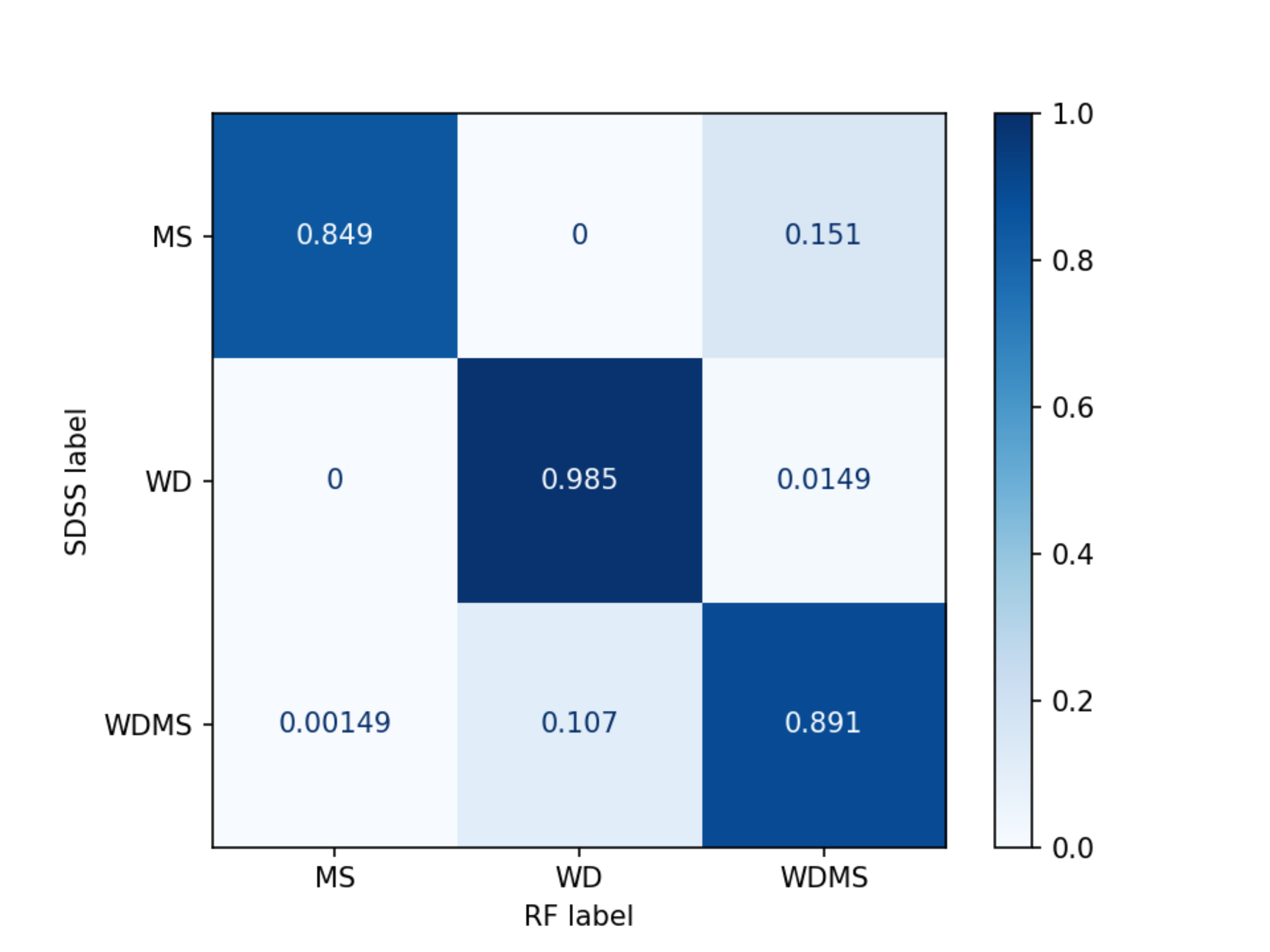}
        \includegraphics[width=0.9\columnwidth,trim=5 0 10 0, clip]{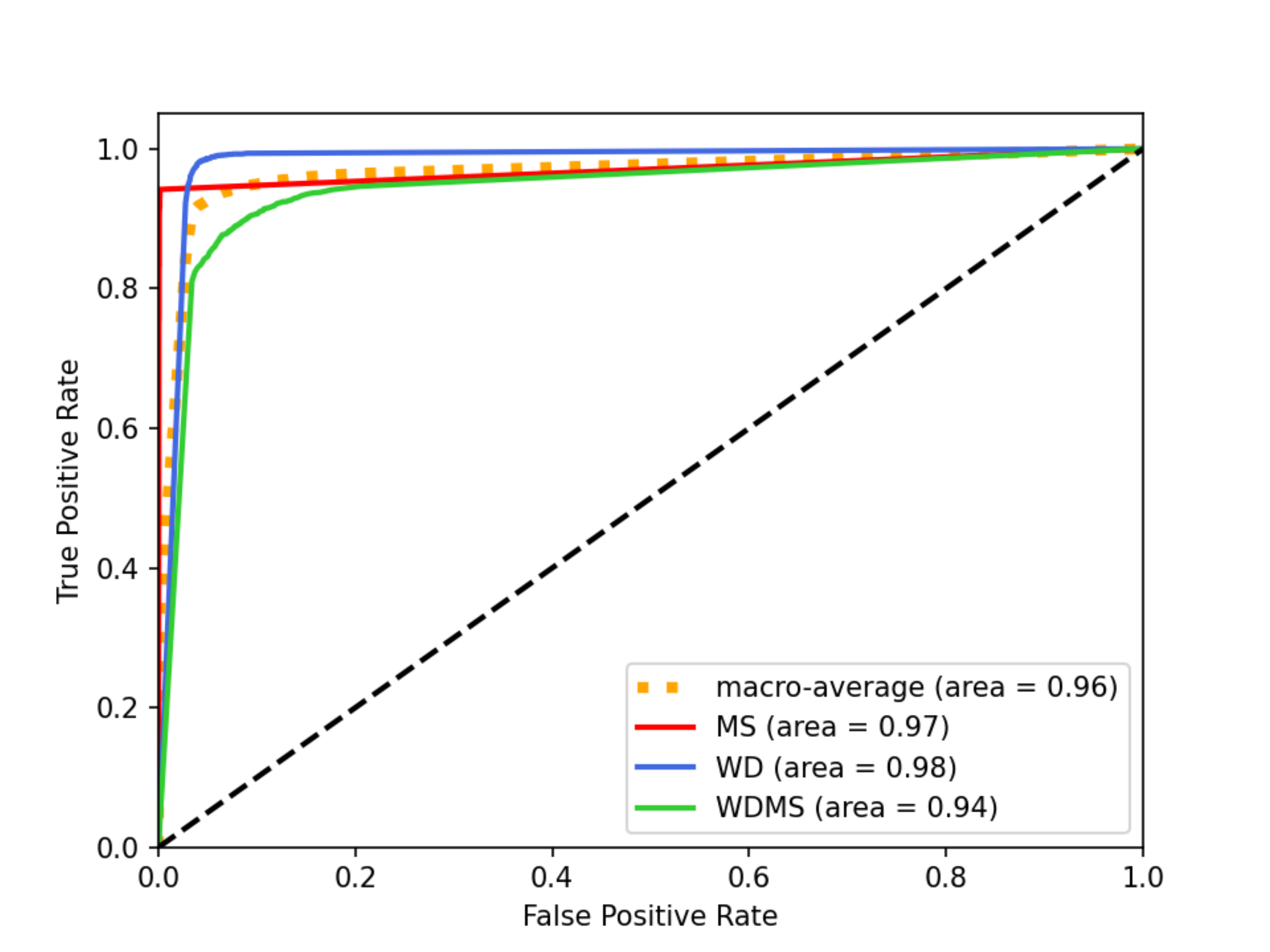}
                        \caption{Confusion matrix (left panel) and ROC curves (right panel) of the SDSS spectra set for the initial Random Forest model.}
    \label{f:class}
\end{figure*}

\begin{figure*}
\centering
    \includegraphics[width=0.9\columnwidth,trim=0 0 10 0, clip]{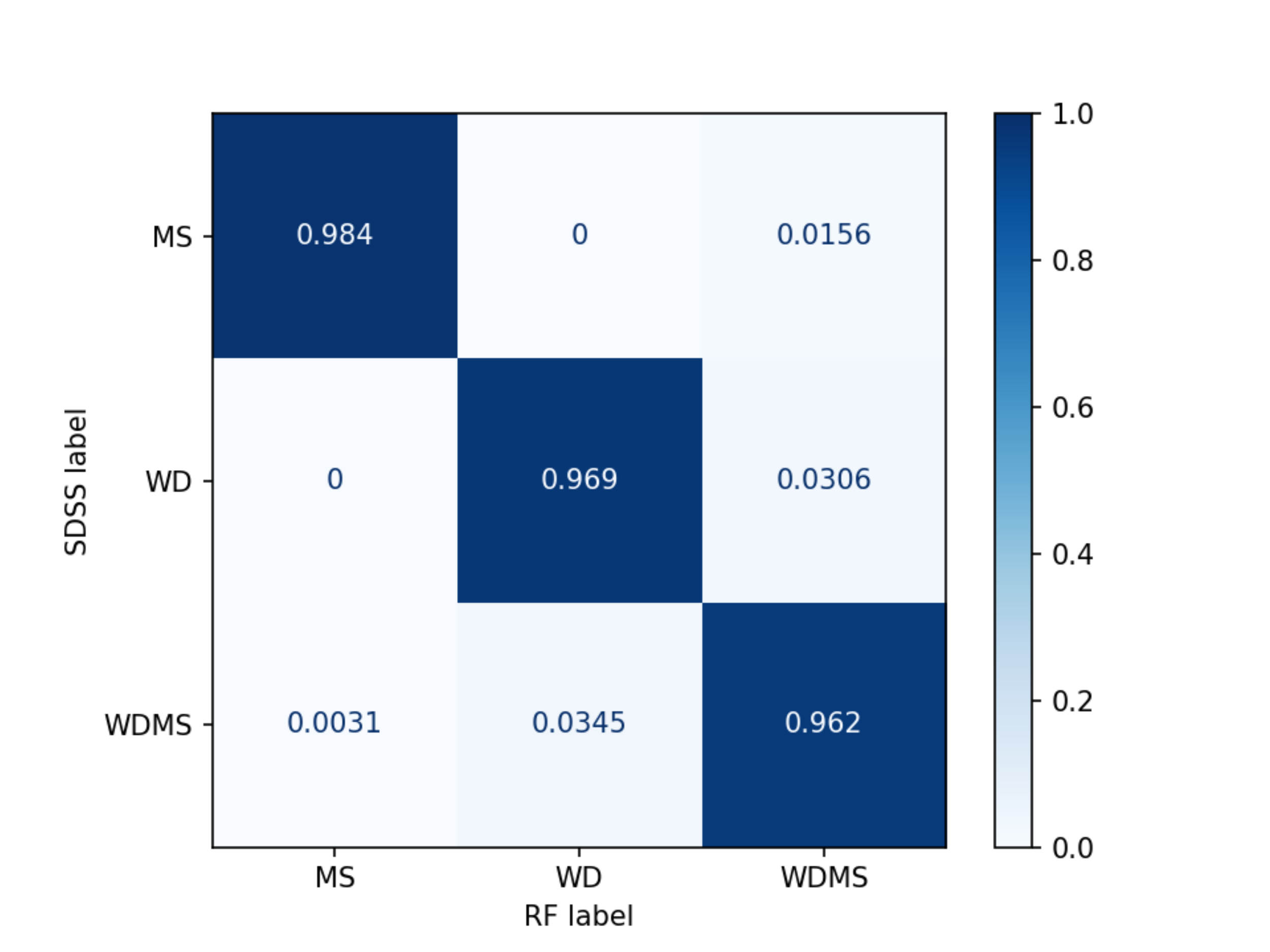}
        \includegraphics[width=0.9\columnwidth,trim=5 0 10 0, clip]{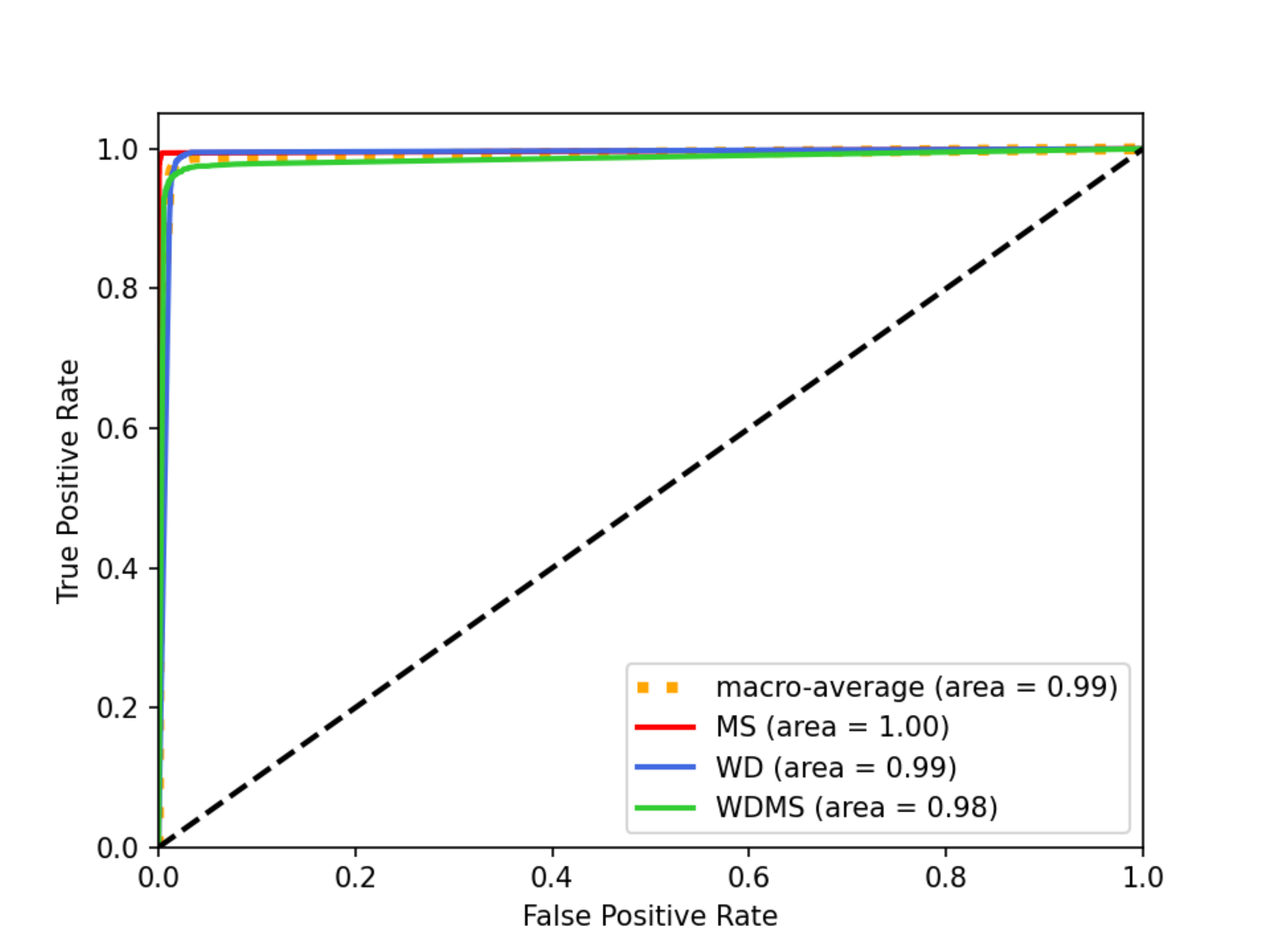}
                        \caption{Same as Fig.\ref{f:class} but for the final Random Forest model. As shown, the agreement between our Random Forest algorithm and the SDSS classification is almost perfect.}
    \label{f:eg_spec}
\end{figure*}

The Random Forest algorithm is set with the adopted default parameters applied in sections \ref{ss:general} and \ref{ss:specific}.  Averaged values are obtained through 100 Random Forest predictions, thus minimizing deviated classifications and allowing us to compute the probability estimates needed for the ROC curves.

The first metric to be analysed is the confusion matrix shown in Figure \ref{f:class}. The three rows indicate the label  assigned by the SDSS spectra, while in the columns we find the Random Forest predicted label.  The WD population shows a nearly perfect agreement between the Random Forest and the SDSS classification. The WDMS has also a higher recall, almost a 90\% with most objects being miss-classified as WD. Finally, the MS population is the one that presents the lowest agreement with the SDSS labels, with almost 15\% of the objects assigned to the WDMS binary population. 

It is important to remark here that the SDSS spectra classified as WDMS binaries clearly display in the vast majority of cases the two components and, as a consequence, they are undoubtedly WDMS stars. Thus we can assume that these objects are correctly classified by SDSS and take them as a reference. We expect that our algorithm can also distinguish the two components. However, there are around 10\% of the objects miss-classified mainly as WDs.

Complementary to the accuracy metric, in the right panel of Fig. \ref{f:class} we show the ROC curve for the three populations under study and the corresponding macro-average curve. For the MS population (red line), we find that with a low threshold   we immediately get a very high recall. This means that the objects that have been classified nearly 100\% of the times as MS, will surely coincide with the SDSS label. For the WD population (blue line) the algorithm and the SDSS label agree in all cases, however, the Random Rorest classifies some WDMS binaries as single WDs. Lastly, the WDMS (green line) curve has the worst performance, and only reaches a recall of around 90\% but with a high FPR. That is, if we want the Random Forest to correctly identified 90\% of WDMS binaries, we must give up on almost 25\% of other objects that will be miss-classified as WDMS.

\subsection{Classification improvements}
\label{ss:cla_imp}

So far, the overall performance of the Random Forest algorithm can be considered as notably good, with an accuracy of 0.904. In order to achieve a higher accuracy and thus a better agreement between the SDSS labels and those provided by the Random Forest, we performed an analysis of the hyperparameters and tested different configurations of the Random Forest. A summary of the modifications of the initial model is detailed as follows:
\begin{itemize}
    \item[i)] \emph{Probability estimates threshold}. We introduced a 99\% threshold for the MS population, and a 95\% for both WDs and WDMS systems. Setting a threshold of the probability estimates for each population improves the  classification at the expense that some spectra remain unlabelled. From the $6\,923$ spectra that were in the test set,  $6\,126$ have been classified -- almost 800 do not accomplish the threshold condition. From those that remained unlabelled, 487 were MS stars, 60 were WDs and 250 were WDMS.
    \item[ii)] \emph{Limiting the stellar parameter range}. Our simulated spectra cover all possible values of effective temperatures as defined from our model grids (see Table\,\ref{t:models}). However, that is not the case for the observed WDMS SDSS sample since it is heavily affected by selection effects \citep{Rebassa2010}. We thus limited our synthetic sample to the range of values covered by the observed SDSS WDMS population, that is, effective temperatures from $2\,500$ to $3\,800\,$K for MS stars and from $7\,000$ to $80\,000\,$K for WDs.
    \item[iii)] \emph{Noise distribution}. In real SDSS spectra, the noise contribution does not affect all the wavelengths in the same way. Conversely, the noise level at the blue and red edges of the spectra are generally higher due to the throughput of the spectrographs, which have less efficiency  in those regions. Thus, we applied a grey type noise so that its contribution is stronger at the blue and end limits, featuring in this way the observed spectra.

\end{itemize}

\begin{figure*}
\centering
        \includegraphics[width=1.65\columnwidth,trim=5 0 0 0, clip]{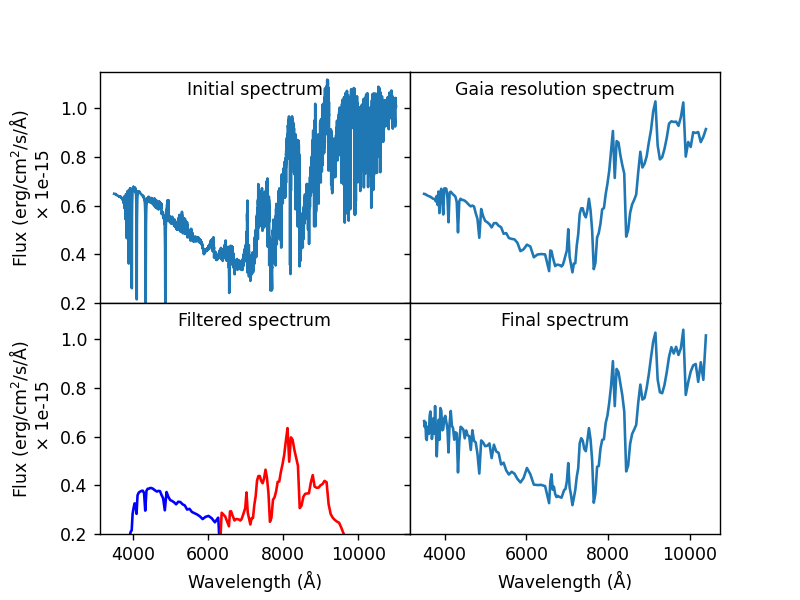}
                        \caption{Example of the building process of a simulated {\it Gaia} WDMS spectrum. The initial noiseless infinite-resolution theoretical spectrum (top left panel) is rescaled to an specific distance of 5 pc and downgraded to the {\it Gaia} resolution (top right panel). It is then filtered through the BP and RP passbands (bottom left panel). A Gaussian noise with a deviation per wavelength interval proportional to the Poisson noise of the number of electrons in that interval is  added and, once defiltered, the final synthetic spectrum (bottom right panel) is obtained. See text for details. }
    \label{f:spectra_process}
\end{figure*}

In the left panel of Figure \ref{f:eg_spec} we present the confusion matrix obtained when the previous improvements have been taken into account in our Random Forest algorithm. The overall accuracy increases to 0.972 (close to a perfect agreement with the SDSS classification labels),  with very low off-diagonal percentages (the maximum is 3.5\%). Due to the limiting parameter range, the MS stars reach a 98\% recall, and the new noise distribution allows the WDMS objects to be retrieved 96\% of the times. All the populations are over the $2\sigma$ confidence level, reassuring the validity of the Random Forest algorithm. The corresponding ROC curves are shown in the right panel of Fig. \ref{f:eg_spec}. The MS population (red line) reaches the perfect case, while the others have areas around 0.98 and 0.99 for the WDMS (green line) and WDs (blue line), respectively.

Of particular interest among the few disagreements are 75 spectra classified as WDMS by the Random Forest that are previously labelled as single WDs (40) or as single MS stars (35). The obvious question is then whether these spectra belong to true single WDs and MS stars or they are WDMS binaries dominated by the flux of one of its components (as indicated by the Random Forest label). A visual inspection of the 75 SDSS spectra  \footnote{They are accessible at: \url{http://wdmsrf.epizy.com/}}  reveals that, in few cases, the two components can actually be discerned. For the remaining ones we do not have sufficient information at hand that robustly allows us to confirm or disprove either hypothesis. Follow-up observations, e.g. for testing radial velocity variations, would be desired for these objects to test their possible WDMS binary nature.

The perfect agreement achieved, after some improvements introduced in  our Random Forest algorithm, leads us to conclude that the artificial classification model performance can be totally equivalent to the previous classification methods applied to SDSS spectra which require a certain level of human expert visualization. 

\section{{\it Gaia} DR3 white dwarf-main sequence spectra classification}
\label{s:gaia}

In this section we study the feasibility of our Random Forest algorithm to classify {\it Gaia} WDMS binary spectra. To that end we first generate synthetic spectra featuring the {\it Gaia} performance for WD, MS and WDMS binary systems. Once our Random Forest is trained, it is then applied to a simulated population of single and binary WD and MS stars within 100\,pc from the Sun. This exercise will give us an idea of how reliable the Random Forest classification applied to real spectra obtained by {\it Gaia} can be, a goal that we will be addressed in a forthcoming publication. 

\begin{figure*}
\centering
    \includegraphics[width=0.39\textwidth,trim=5 0 0 0, clip]{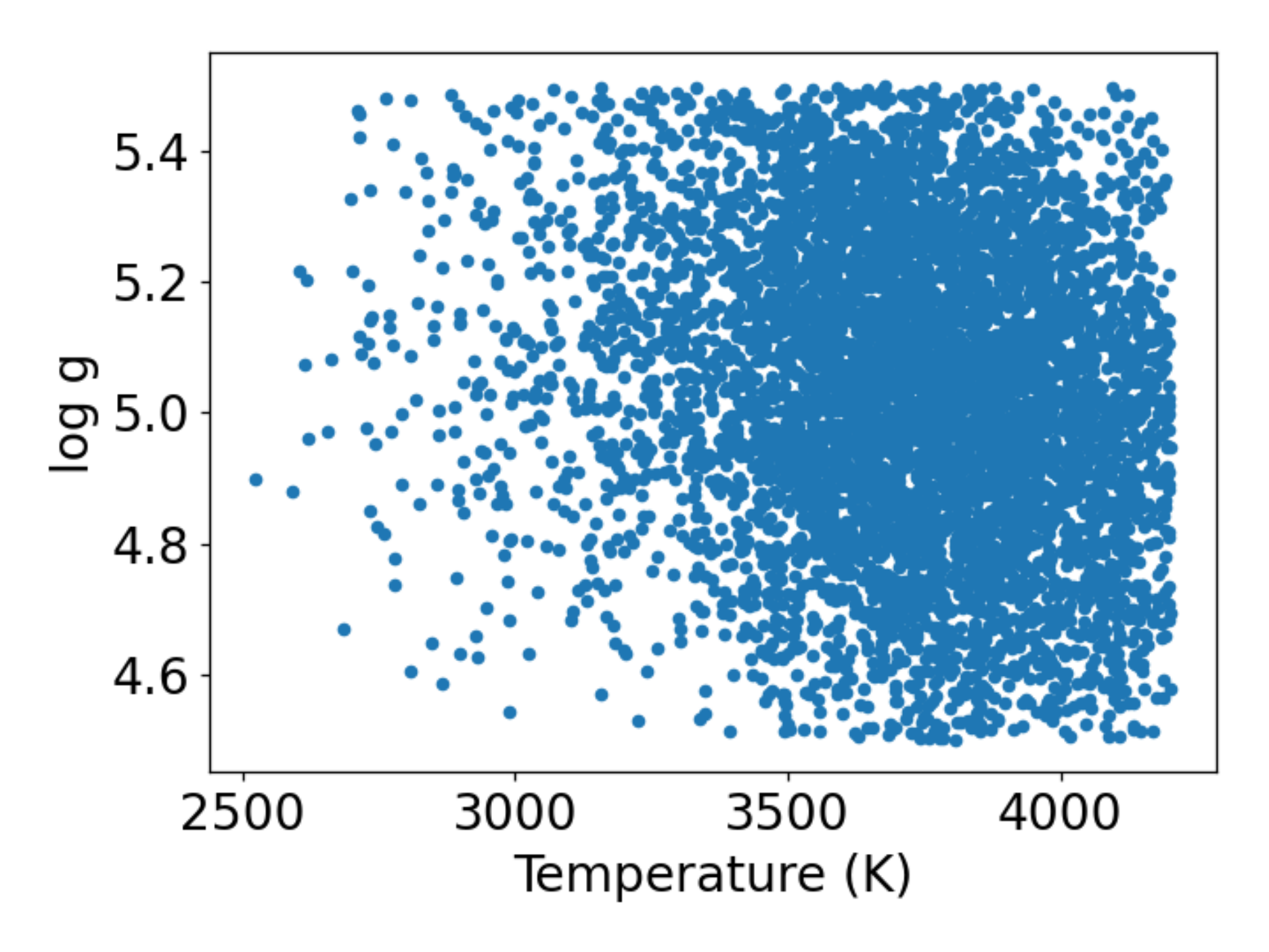}
    \includegraphics[width=0.39\textwidth,trim=5 0 0 0, clip]{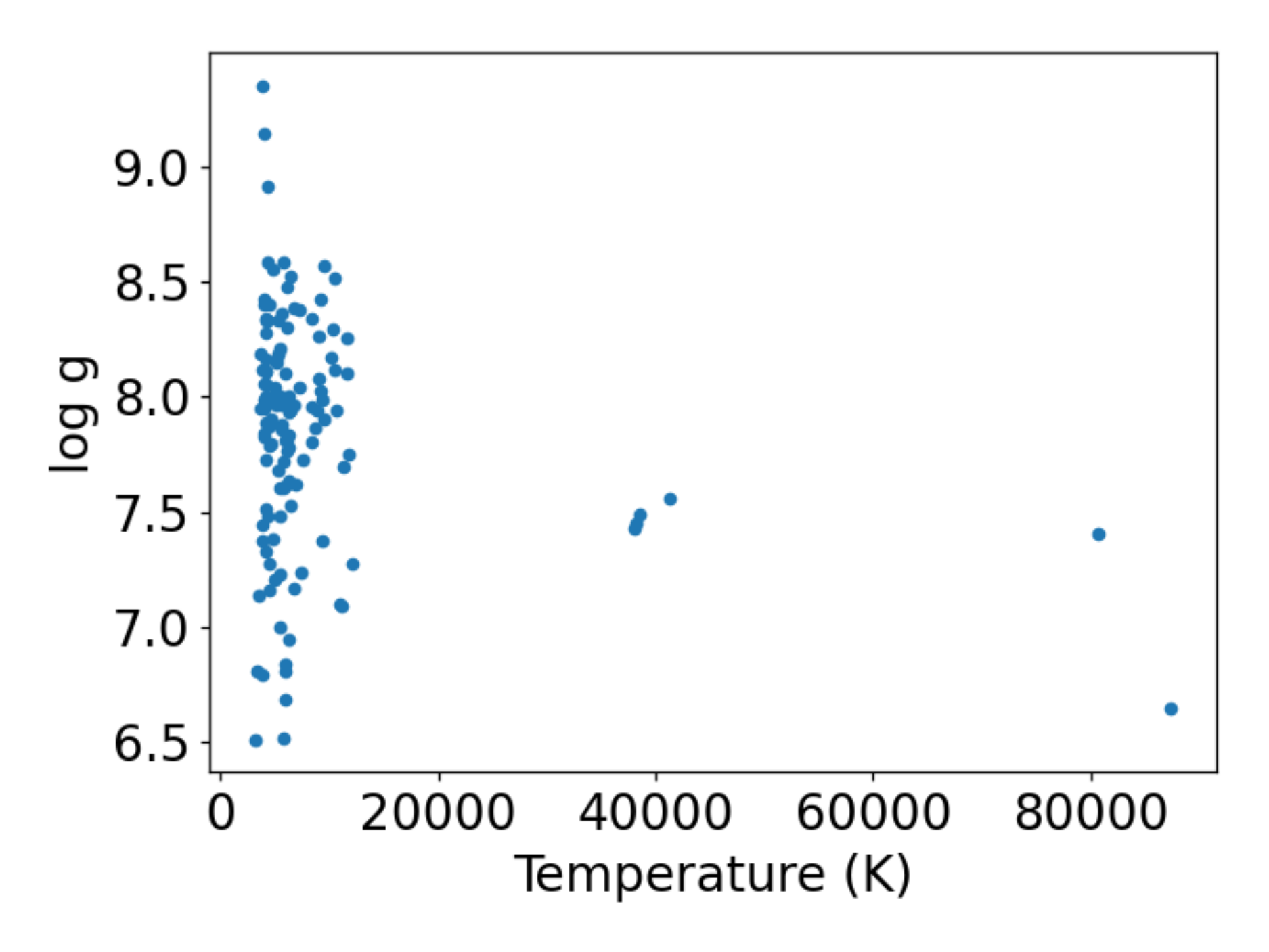}
    \includegraphics[width=0.39\textwidth,trim=5 0 0 0, clip]{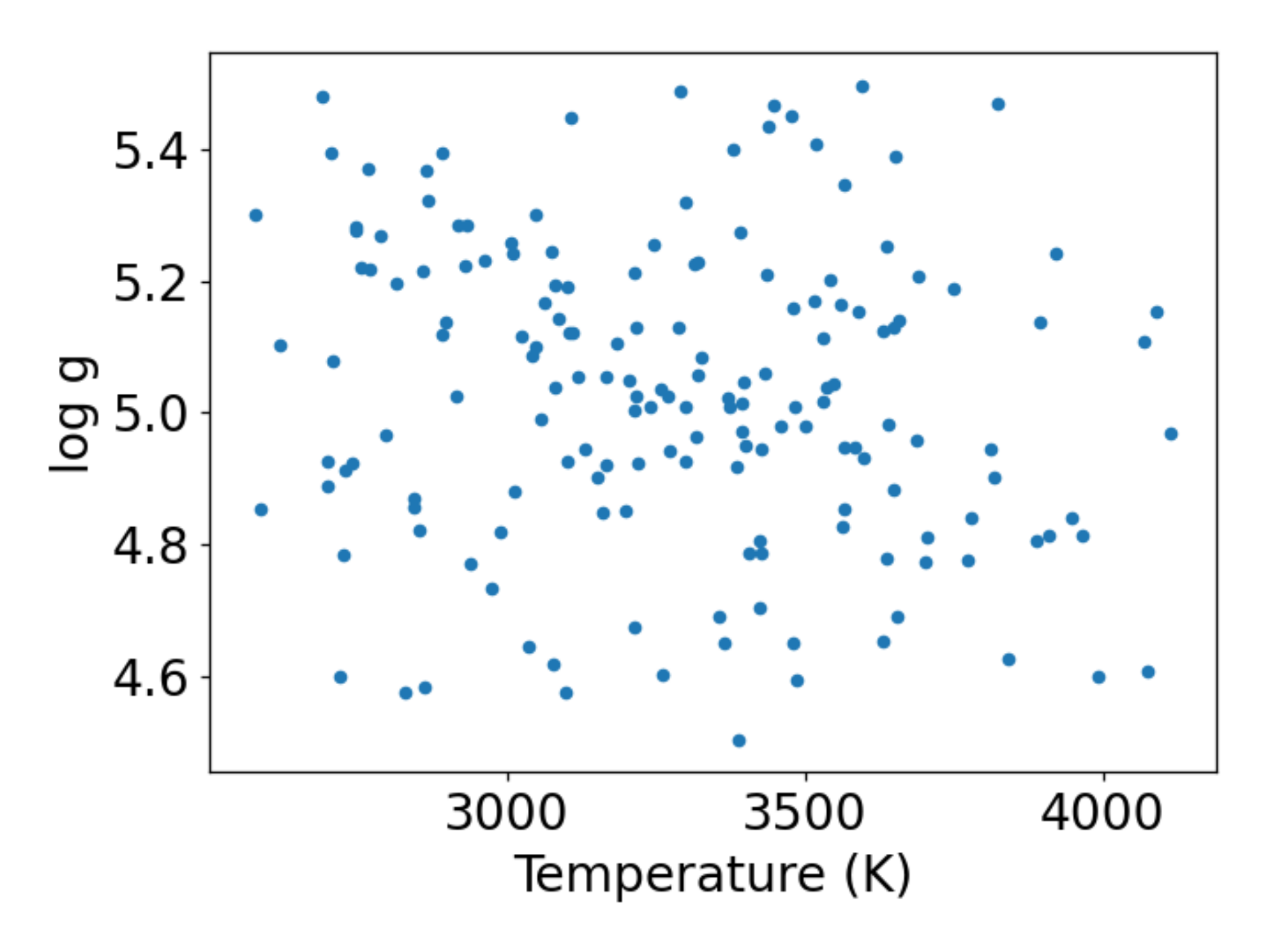}
    \includegraphics[width=0.39\textwidth,trim=5 0 0 0, clip]{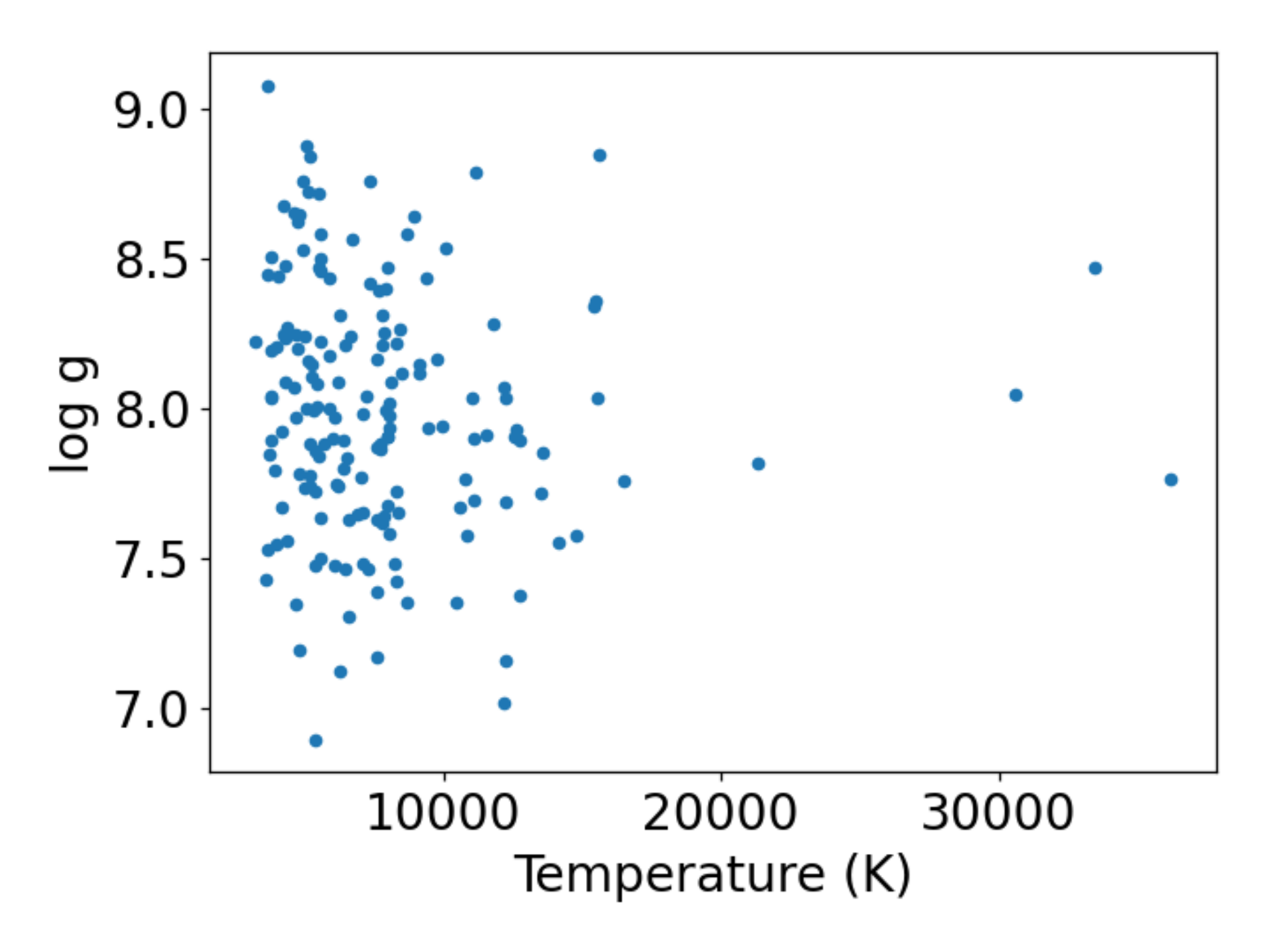}
    \caption{Distribution of stellar parameters of the {\it Gaia} accessible population of single MS stars (top left), single WDs (top right), MS star components in WDMS (bottom left) and WD components in WDMS (bottom right).}
    \label{fig:WDandWDMSpar}
\end{figure*}

\subsection{\emph{Gaia} DR3 synthetic spectra}
\label{ss:synspe}

Our first task is to create a set of synthetic spectra emulating the {\it Gaia} performance. As it is well known, the {\it Gaia} spectrograph consists of two low-resolution prisms, one for the blue wavelengths (Blue Photometer or BP) from $3\,300$ to $6\,800\,$\AA\, and the other for the red wavelengths (Red Photometer or RP), between $6\,400$ and $10\,000\,$\AA\, \citep{Gaia2016}. The expected resolving power is relatively low  and not constant, being it a function of the wavelength \citep[see Figure 3 from][]{Carrasco2021}. The mean $R$ is around 66 and the expected SNR of the spectra depends on the stellar object and its magnitude. SNR estimates for the MS and WD stars with a magnitude of 13 are shown in Table\,\ref{tab:gaia_SNR}. 

\begin{table}
    \caption[]{Expected SNR for {\it Gaia} WD and MS star spectra for an apparent magnitude of $G=13$ mag (Carrasco private communication).}
    \label{tab:gaia_SNR}
         \begin{center}

    \begin{tabular}{lcc}
            \noalign{\smallskip}
            \hline
            \noalign{\smallskip}
               & SNR BP & SNR RP \\
            \noalign{\smallskip}
            \hline
            \noalign{\smallskip}
        MS star & 278 & 1036.2 \\
            \noalign{\smallskip}
            \hline
            \noalign{\smallskip}
        WD star & 451.9 & 487.2 \\
            \noalign{\smallskip}
            \hline
    \end{tabular}
\end{center}
\end{table}

In Figure \ref{f:spectra_process} we show the process that leads from our theoretical initial model to our final {\it Gaia} simulated spectrum (for a mathematical description of the process see Appendix \ref{a:noisemodel}). As an illustrative example we take a WDMS system located at distance $d=5\,$pc from the Sun, with $G=16$ mag  that implies an associated spectrum with ${\rm SNR_{RP}}=35$  and ${\rm SNR_{BP}}=18$. The initial noiseless and virtually infinite resolution spectrum is plotted on the top left panel of Fig.  \ref{f:spectra_process}. Firstly, as the initial models are computed at a fixed distance of $200\,$pc, we rescaled the spectrum  to the particular distance of the system. Then, the spectrum is downgraded according to the {\it Gaia} nominal spectral resolution relationship from \citep{Carrasco2021}. The resulting  spectrum is shown on the top right panel. In the next step the spectrum is filtered through the BP and RP passbands. In order to avoid filter overlaps and without losing generalization in our procedure, we consider that filters BP and RP, end and start, respectively, at $6 350\,$\AA.  The filtered spectrum is plotted in the left bottom panel of Fig. \ref{f:spectra_process}. 

Finally, we add noise to our filtered spectrum. As previously stated, the estimated SNR for each filter and kind of object (WD or MS) for a fixed $G$ magnitude of 13 mag is presented in Table \ref{tab:gaia_SNR}. For a different $G$ magnitude the SNR is computed assuming that it  is proportional to the flux. In case of binary systems, the corresponding WD and MS star fluxes are added and the SNR is recalculated to the corresponding $G$ magnitude of the system (see equations \ref{e:SNRG} and \ref{e:SNR1SNR2} from Appendix \ref{a:noisemodel}). According to {\it Gaia} performance, it can be assumed that Poisson noise dominates for faint sources (typically $G>12$ mag; as it is the case of WDMS systems), while CCD readout noise can be disregarded. Hence, the filtered spectrum, which is provided in flux units, is converted to number of electrons per wavelength interval. That way a Poissonian noise proportional to the number of electrons can be naturally introduced in the modelling. In this process we assume that the conversion factor from photons to electrons is constant all along the wavelength range of the filter. Additionally, the specific characteristics of the CCD device are engulfed in a normalization constant. Thus, to each wavelength interval of the filtered spectrum we add a Gaussian noise whose deviation follows a Poissonian noise, that is, it is proportional to the square root of the number of electrons in that interval. The normalization constant is calculated in such a way that the full spectrum has the SNR previously derived for that system. The final resulting spectrum once the noise has been added and then defiltered is shown in the bottom right panel of Fig. \ref{f:spectra_process}. As can be observed, it results in a realistic gray noise model of the {\it Gaia} spectrum.

\subsection{{\it Gaia} white dwarf-main sequence synthetic population}
\label{ss:synpop}

Once WD, MS and WDMS {\it Gaia} spectra are modeled, we take advantage of a detailed population synthesis modeling of the Galaxy widely used in the analysis of the WD population \citep[see][and references therein]{Torres2019,Torres2022}. The simulator, based on Monte Carlo techniques, generates the expected stellar parameters of the population of single WDs, single MS stars and WDMS binaries in the Galaxy that can be accessible by the {\it Gaia} satellite. For the analysis presented here we adopt  a standard model (a full description of the model, named as Model 1, can be found in \citealt{Torres2022}). The physical inputs consist in a flat initial-mass-relation distribution, $n(q)\propto 1$, an inverse initial orbital separation, $f(a)\propto a^{-1}$, and a binary fraction of $f_{b}=0.50$. The synthetic population is mixed with a ratio 74:25:1 for the thin and thick disk, and halo Galactic components, respectively, within $100\,$pc from the Sun. Photometric and astrometric errors have also been introduced for each object of the simulated sample according to {\it Gaia}’s performance. 

\begin{figure*}
\centering
    \includegraphics[width=0.33\textwidth,trim=25 0 25 10, clip]{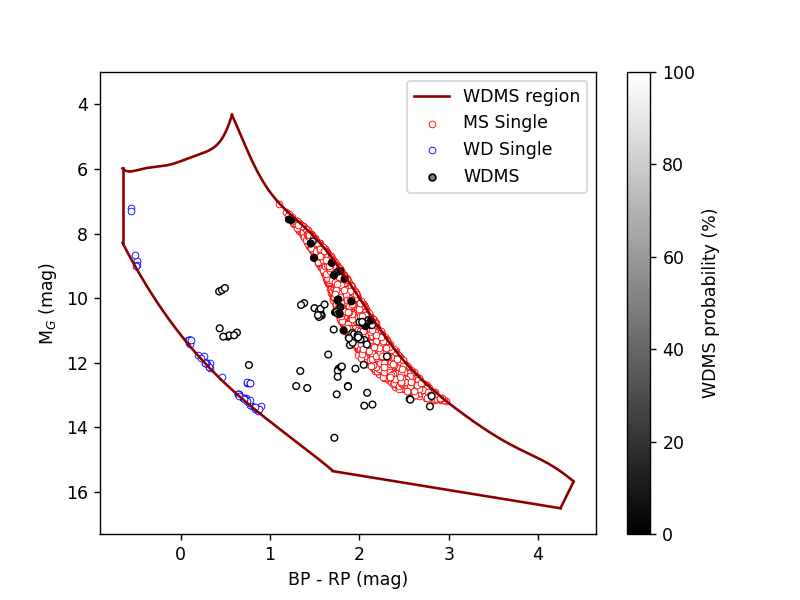}
    \includegraphics[width=0.33\textwidth,trim=25 0 25 0, clip]{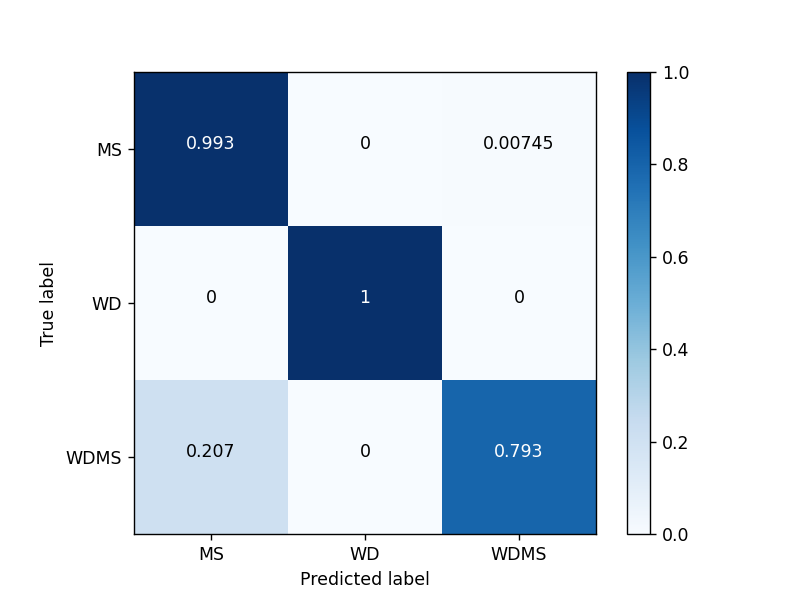}
       \includegraphics[width=0.33\textwidth,trim=25 0 25 0, clip]{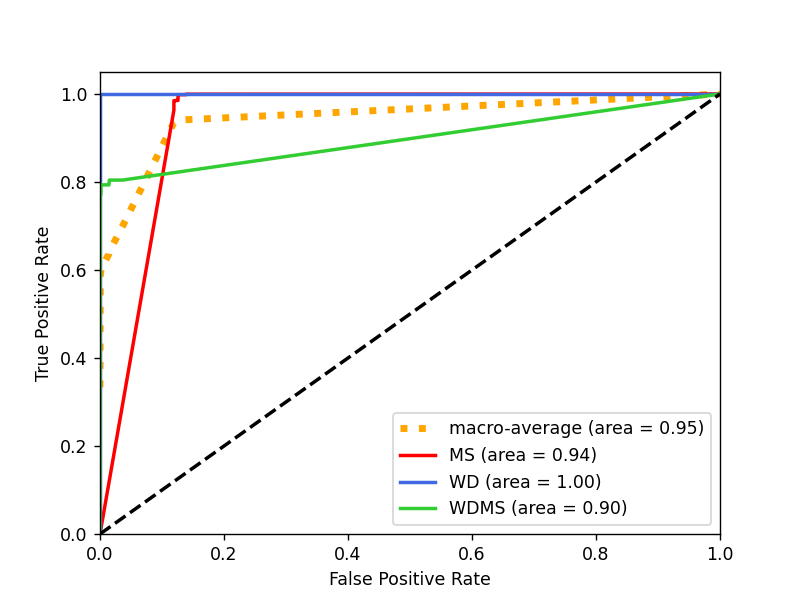}
    \includegraphics[width=0.33\textwidth,trim=25 0 25 10, clip]{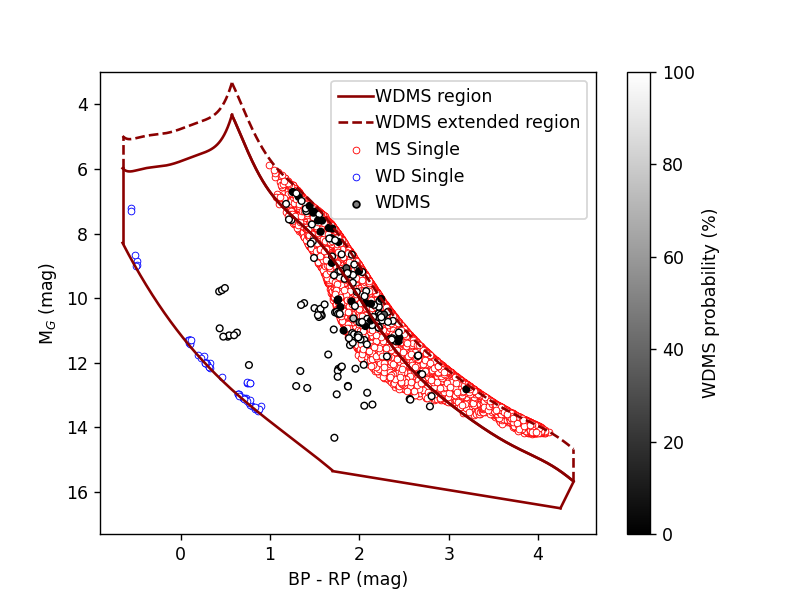}
    \includegraphics[width=0.33\textwidth,trim=25 0 25 0, clip]{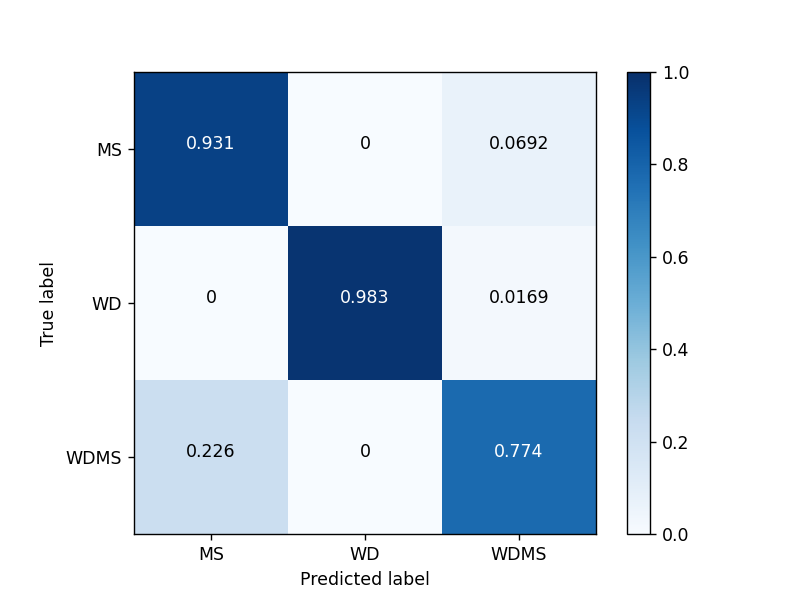}
       \includegraphics[width=0.33\textwidth,trim=25 0 25 0, clip]{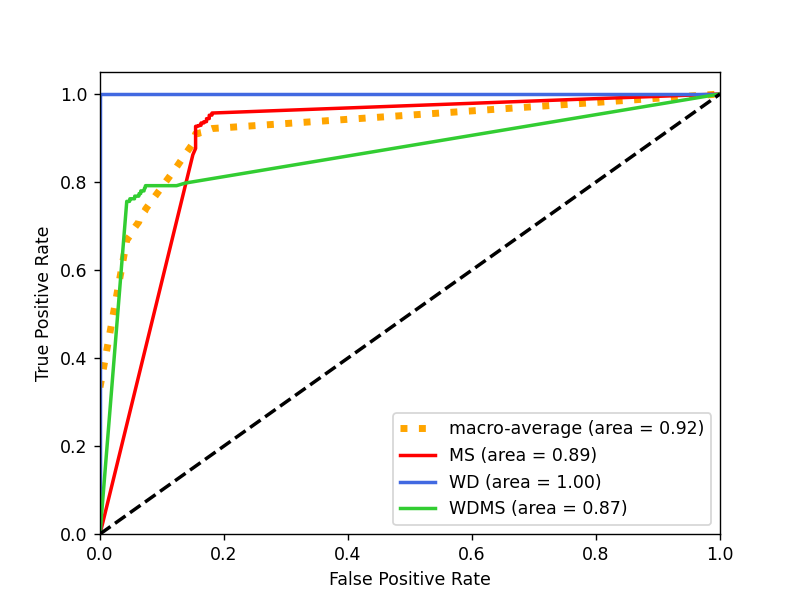}
    \caption{Top panels: HR-diagram (left panel) of WDs (blue open circles), MS stars (red open circles) and  WDMS systems (grey solid circles) for those spectra with $G<18$ accessible by {\it Gaia} within the WDMS region (red solid line) defined by \citet{Rebassa2021}. The performance of our Random Forest in classifying these population is shown in the  corresponding confusion matrix (middle panel) and the ROC curve (right panel). Bottom panels: the same as previous panels when the region defined by \citet{Rebassa2021} is extended 1 mag up into the MS region (red dashed line). Even in those extreme conditions the algorithm is able to identify $\sim$60\% of WDMS binary systems within the extended region (see text for details). }
    \label{fig:Gaiaresults}
\end{figure*}

In Figure \ref{fig:WDandWDMSpar} we display the distribution of stellar parameters, namely their effective temperatures and surface gravities, for the different synthetic populations generated by our simulator. The result shows that the percentages of the different populations are clearly unbalanced, resulting in 5992:126:167, i.e. 95.34\% of single MS stars, 2\% of WDs and 2.66\% of binaries. Based on this simulation we build the testing sets by applying a bivariate interpolation of the temperatures and surface gravities on the three spectral grids of MS, WDs and WDMS (see Section \ref{ss:synsamp}.), thus obtaining the corresponding synthetic spectra for each simulated star of the sample following the procedure described in the previous section. 

\subsection{Classification of  {\it Gaia} white dwarf-main sequence systems}
\label{ss:class}

In Figure \ref{fig:Gaiaresults} we show the results of our Random Forest algorithm when applied to the simulated spectra of the {\it Gaia} 100 pc synthetic population of WDs, MS stars and WDMS systems. Following {\it Gaia} performance, we have selected only those spectra with $G<18$ mag. Moreover, as a starting point to test our algorithm, we used the region within the HR diagram (solid red line in left panels of Figure \ref{fig:Gaiaresults}) defined by \cite{Rebassa2021} to select unresolved WDMS. Within this region a certain level of contamination from single low-mass WDs and especially from single low-metallicity MS stars is expected as shown by our simulations (blue open circles and red open circles, respectively in Fig. \ref{fig:Gaiaresults}). However, as revealed by the confusion matrix on the top middle panel of Fig. \ref{fig:Gaiaresults}, the performance of the Random Forest can be considered as excellent,  achieving a practically perfect accuracy of 0.99. This is due to the fact that the sample within our selected region is highly unbalanced, representing the MS star population 97.6\% of the whole sample, while WDs and WDMS systems represent only 1 and 1.4\%, respectively, of the objects. Even so, practically all single WDs and MS stars and nearly 80\% of the WDMS systems are correctly classified. Only in the region where WDMS and single low-metallicity MS stars overlap, WDMS binaries are miss-classified as MS stars (see gray scale on top left  panel of Fig. \ref{fig:Gaiaresults}) and, at the same time, some MS stars are erroneously assigned as  WDMS systems. The performance of the algorithm for each sub-population can be analyzed in the ROC curves illustrated in the top right panel of Fig. \ref{fig:Gaiaresults}. As shown, the performance of the Random Forest algorithm for correctly classifying single WDs is perfect, as it is the case for MS stars when adopting a FPR threshold larger than $\approx0.17$. The performance of the algorithm in classifying WDMS is also notably good, with a TPR always above 0.80 and a total ROC area of 0.90.

However, it is important to emphasise that only 9\% of the total underlying population of unresolved WDMS binaries falls within the region defined by \cite{Rebassa2021}, while the remaining $91\%$  lay above this region  (that is the area where single MS stars of solar metallicity are located). From the point of view of the spectral classification, we have to face two problems in this region. On one hand, the large quantity of MS stars. And, on the other hand, the SEDs of  WDMS systems which overlap in the HR diagram with single MS stars are normally dominated by the flux of the MS companion, being this the reason why the vast majority of such WDMS binaries remain observationally undetected. Nevertheless, motivated by the excellent results achieved by of our Random Forest algorithm, we extend our analysis to test whether or not it can help in identifying these elusive binaries. For that purpose we extend 1 magnitude up the upper border of the WDMS region defined by \citet{Rebassa2021}. The new selection region is marked by a dashed-red line in the bottom left panel of Fig.\ref{fig:Gaiaresults}.   As shown, a large number of WDMS, as well as MS stars, are now included in the extended region. The sample is now totally dominated by MS stars (99\%), while WDs and WDMS binaries are just 0.3 and 0.7\% of the  sample. The corresponding confusion matrix is shown in the bottom middle panel of Fig. \ref{fig:Gaiaresults}. The performance of the algorithm, even slightly worse that in the previous case, can still be considered as notably good. Practically all WDs, and nearly 93\% of the MS stars are correctly classified. Regarding WDMS binary systems, 77\% are properly identified, being those miss-classified assigned to MS stars. It is important to remark that in this extended region due to the large quantity of MS stars the identification of WDMS systems through their SEDs is extremely difficult. However, the Random Forest algorithm is able to correctly identify nearly 60\% of WDMS binary systems. However, the drawback is that even though the percentage of MS stars miss-classified as WDMS is small ($\approx 7\%$), that represents a decrease in the number of TPR of the WDMS population, as shown in the ROC curves (bottom right panel of Fig.\,\ref{fig:Gaiaresults}). The performance of the algorithm in correctly classifying WDMS is reduced to a ROC area of 0.87, which is also related to a slight decrease in the TPR of the MS star population. 

We conclude that, even with the intrinsic difficulties for disentangling WDMS binaries from single MS stars in the extended region, our results demonstrate that a classification of {\it Gaia} WDMS spectra based on a Random Forest algorithm is feasible and can be considered as a powerful tool for identifying binary WDMS systems, even those that so far have remained elusive.

\section{Conclusions}
\label{s:conclusions}

We have proven that a Random Forest algorithm is an optimal tool for classifying single WDs or MS stars, as well as WDMS binaries among large spectroscopic databases such as the SDSS or {\it Gaia}\. We first analyzed the capabilities of the algorithm by classifying synthetic spectra for the full space-parameter range of effective temperatures and gravities of WDs and MS stars. The synthetic spectra were generated combining the individual WD and MS spectra, which cover a wide range of temperatures and surface gravities. Our results show that the accuracy of the classification depends only on the SNR of the spectra for resolving powers $R$ above $\approx$300. For a typical case with a resolving power of $1\,800$ and SNR=10 the algorithm is able to recover practically all MS stars and WDs, 93 and 99.3\%, respectively, while up to 74\% of WDMS systems can be correctly classified.

The Random Forest algorithm was also tested using the SDSS catalogue where nearly $7\,000$ selected WD, MS and WDMS stars were already labelled. The agreement between our algorithm and the SDSS classification was practically total, achieving 97\% of match. Taking into account that, in our previous theoretical analysis we found a recall ratio of WDMS systems of 74\% by our algorithm, we can hypothesize that a similar ratio is presented in the SDSS labeling. This would imply that not all WDMS systems are properly recovered by the SDSS classification.  Among the discrepancies, we found 75 spectra classified by the Random Forest as WDMS that were initially labelled by SDSS as WDs (40) or MS stars (35). Visual inspection of these spectra reveals that, for some of them, the two components can actually be discerned, while for others  further follow-up observations are required to confirm whether or not they are single stars or binaries. This result gives indications that the Random Forest algorithm is capable to some extent to correctly classify WDMS binaries even in the case that the flux is dominated by one of their components.

Once our algorithm has been proven to be a useful tool for classifying WDMS spectra from actual surveys, we explored the possibilities of applying it to a synthetic sample of spectra mimicking {\it Gaia} observations. First, we introduced a realistic modeling of {\it Gaia} spectral noise taking into account, among other things the SNR as a function of the G magnitude for different sources such as WDs or MS stars, as well as the error that arises during the filtering through the {\it Gaia} BR and RP passbands. Then, with the aid of a detailed Monte Carlo simulator of the Galactic WD population, we generated the parameter space for WDs, MS stars and WDMS systems observable by {\it Gaia} within 100 pc from the Sun. For each object of the simulated sample we assigned a synthetic \emph{Gaia} spectrum with their corresponding SNRs. We chose as our initial area for evaluating the Random Forest algorithm the region within the HR-diagram defined by \cite{Rebassa2021} for selecting unresolved WDMS binary objects. The performance of the algorithm within this region is notably good, being able to identify nearly 80\% of WDMS systems. The analysis  has been then extended to regions of the HR-diagram where the presence of MS stars is dominant and, at the same time, WDMS systems are practically overshined by the MS components. Even though from an observational point of view the detection of WDMS in these areas is extremely difficult, the algorithm  was able to correctly identify nearly 60\% of the WDMS systems in that extended region, which further supports the potential of the Random Forest to identify these elusive objects.

Along this work we have shown that the Random Forest algorithm is a powerful tool for classifying large samples of spectra, specifically those of WDs, MS stars and WDMS binary systems. In particular, once provided with a realistic modeling of spectral noise and the population parameters, we have proven that the Random Forest is a feasible tool for identifying and correctly classifying WD, MS and WDMS systems observed by {\it Gaia}.

\begin{acknowledgements}
This work was partially supported by the MINECO grant  PID2020-117252GB-I00. ARM acknowledges support from grant RYC-2016-20254 funded by MCIN/AEI/10.13039/501100011033 and by ESF Investing in your future. We would also like to acknowledge the fruitful information provided by M. Carrasco and J. de Bruijne.
\end{acknowledgements}


\bibliographystyle{aa}
\bibliography{RFSP}


\appendix

\section{Random Forest: hyperparameters, metrics and scoring}
\label{a:hyper}

\subsection{Hyperparameters}

For the present project we have used the scikit-learn\footnote{\url{https://scikit-learn.org}} library for Python, which already includes a Random Forest classifier \citep{Pedregosa2011}.
Some of the most relevant hyperparameters are defined  and in Table \ref{tab:valueshyper} a list of the adopted values is presented. 
\begin{itemize}
    \item \textbf{Number of estimators}: the number of Decision Trees that form the Random Forest.
    \item \textbf{Criterion}: the function that is used in order to calculate the quality of the split. Generally, Gini index or the entropy index are used. In our case, Gini impurity index has been adopted, defined as: 

\begin{equation}\label{E:giniimpurity}
    I_{\rm Gini}(j) = \sum_j{p(i|j) \cdot [1-p(i|j)]}
\end{equation}

where $(p(i|j)$ is the probability of picking an element of class $i$ at certain node $j$, so it will be 0 when all the samples of a branch are in the same class and 0.5 when the mixture has the same elements of each class. 

    \item \textbf{Minimum number of samples required to split a node}: if there are more samples than this value, the algorithm will continue the splitting process; if not, the current node will be considered a Leaf Node.
    \item \textbf{Minimum number of samples required to be at a Leaf Node}: the current node will be considered to be a Leaf Node if there are at least a number of samples equal to this value in the node. 
    \item \textbf{Maximum depth}: the algorithm finishes a Decision Tree when there are no more mixed subsets with different classes, when the number of samples is less than the minimum number of samples required to split a node, or when the tree reaches the maximum number of Decision Nodes set by this hyperparameter. 
    \item \textbf{Maximum number of features to consider when looking for the best split}: the  number of random features that the algorithm takes into account when doing a split can be controlled with this parameter. 
    \item \textbf{Maximum samples to draw from the training set to train each base estimator}: if bootstrap is enabled, this value sets the number of samples used in order to build each estimator. If there is no set value, the algorithm will use all the samples.
    \item \textbf{Class weights:} parameter that allows to set weights to the different classes. This is useful in imbalanced datasets where there are many objects of one population and few of another, as it biases the calculation of the Gini index in favour of the minority class at the cost of allowing some errors in the majority one.
\end{itemize}

\begin{table}
    \caption[]{Hyperparameters optimal values adopted in the present work. Default values are marked with asterisk.}
    \label{tab:valueshyper}
\begin{center}
    \begin{tabular}{lc}
            \noalign{\smallskip}
            \hline
            \noalign{\smallskip}
        Hyperparameter &  Optimal  value \\
            \noalign{\smallskip}
            \hline
            \noalign{\smallskip}
        Number of estimators & 200 \\

        Maximum depth & None*  \\

        Minimum samples split & 2*  \\

        Minimum samples leaf & 1*  \\

        Maximum features & 500 \\
            \noalign{\smallskip}
            \hline
    \end{tabular}
\end{center}
\end{table}

\subsection{Metrics and scoring}

Some of the most relevant metrics and scoring used along this work are presented:

    \begin{itemize}

    \item \textbf{Confusion matrix}, $\textbf{M}$: an element of the matrix $M_{i,j}$ corresponds to all the objects of true class $i$ and labelled as $j$. The diagonal indicates the number of objects  that were actually classified according to its true label, while the off-diagonal elements represent the misclassified objects.

    \item \textbf{Accuracy score}: defined as 
      \begin{equation}\label{E:accuracy}
        accuracy = \frac{tr(M)}{N_{\rm total}} ,
    \end{equation}
    
  represents the fraction of correctly classified elements divided by the whole population. Its maximum value  1 indicates that all the elements are correctly classified, while its minimum 0 that all the elements are incorrectly classified.

    \item \textbf{Recall score}: it measures the fraction of correctly labelled as class $i$ from the whole set of objects that actually belong to that class. Defined as
    \begin{equation}\label{E:recall}
        recall_i = \frac{M_{i,i}}{\sum_{j} M_{i,j}},
    \end{equation} 
    
    where the sum term refers to the sum of the elements of the row $i$.
    
    It is also called sensitivity or probability of detection, because it indicates the True Positive samples, that is, the hits of the algorithm for a certain class. In a normalized confusion matrix, the recall scores directly correspond to the diagonal values.
    
    \item \textbf{Precision score}:  indicates the amount of objects correctly classified as $i$ in comparison with the total amount of objects labelled as $i$. This value quantifies the ability of the classifier to not label as class $j$ an element of class $i$, which would be:
    
    \begin{equation}\label{E:precision}
        precision_i = \frac{M_{i,i}}{\sum_{j} M_{j,i}},
    \end{equation}
    
    where the sum term refers to the sum of the elements of the column $i$.
    
    \item \textbf{F1 score}: also named F-score is a weighted average of both, the precision and recall scores.  Its mathematical expression is given by:
    
    \begin{equation}\label{E:f1score}
        {\rm F1}_i = \frac{precision_i \times recall_i}{precision_i + recall_i}.
    \end{equation}

    \item \textbf{ROC curve}: the Receiver Operating Characteristic (ROC) curve, although it is primarily intended for binary classifications, it can be extended to multi-class problems such as the present one. This curve represents the performance of a classifier while varying its discrimination threshold. The resulting points are the True Positive Rate (TPR), also known as recall or probability of detection, versus the False Positive Rate (FPR), also named as probability of false alarm. We could express those three values as:
     
    \begin{equation}\label{E:tpr}
        TPR = \frac{True\:Positives}{Positives} = \frac{True\:Positives}{True\:Positives + False\:Negatives}
    \end{equation}   
    and
    \begin{equation}\label{E:fpr}
        FPR = \frac{False\:Positives}{Negatives} = \frac{False\:Positives}{False\:Positives + True\:Negatives}.
    \end{equation}  
    
    In a multi-class problem, the Positives would be all those elements that actually belong to a class $i$, whether they are correctly classified or not, and the True Positives would be those who are correctly classified as a class $i$. On the other hand, the Negatives would be all the other samples that do not belong to a class $i$, while the False Positives are the elements that do not belong to a class $i$ but are classified in it. To sum up:
    
       \item \textbf{AUC curve}: the Area Under the Receiver Operating Characteristic Curve (ROC AUC), which is the area under the  ROC curve. This value basically summarizes the information of the curve in a number that goes from 0.5 in the worst case, to 1 if the classifier does not make mistakes.

\end{itemize}

\section{\emph{Gaia} spectral noise modeling}
\label{a:noisemodel}

Let $\phi$ be the flux of a noiseless spectrum  measured in units  of erg\,${\rm cm}^{-2}\,{\rm s}^{-1}\,$\AA$^{-1}$. The initial flux, $\phi_0$, computed at distance $d_0$ (in our case, $d_0=200\,$pc) can be directly rescaled at distance $d$ as $\phi_d=\phi_0\left(200/d[pc]\right)^2$.
Taking into account the resolving power, the flux is discretized in $n$ intervals of length $\Delta\lambda$, $\phi_i$, with $i=1\ldots n$.
If the wavelength interval $\Delta\lambda$ is small enough, we can estimate the  number of electrons generated in interval $i$, $E_i$, as a linear function from the number of photons derived from the flux, $\phi_i$ as
\begin{equation}
    E_i=\phi_i\times\frac{\Delta A\cdot\Delta t}{\epsilon_{\rm photon}}\times\Delta\lambda\times f
\end{equation}
where $\Delta A$ is the area covered by the CCD camera, $\Delta t$ the time interval, $\epsilon_{\rm photon}$ the energy of the photon of wavelength $\lambda_i$ and $f$ a conversion function from photons to electrons.  Assuming that $f$ is independent of $\lambda_i$ (or at least for the wavelength range of the spectrum) and recalling that the resolving power is $R=\lambda/\Delta\lambda$, the previous equation leads to:
\begin{equation}
\label{e:numbere}
 E_i=\mathcal{C}\phi_i\lambda_i^2  
\end{equation}
where $\mathcal{C}$ is a constant that encompasses the rest of parameters. 

The signal S and the noise N are defined in the standard way as
\begin{equation}
\label{e:S}
 S=\sqrt{\sum_i^n\phi_i^2}
 \end{equation}
 and
 \begin{equation}
 \label{e:N}
 N=\sqrt{\sum_i^n\varepsilon_{\phi,i}^2}  
\end{equation}
where $\varepsilon_{\phi,i}^2$ is the error associated to the flux, $\phi_i$.

For each interval $i$, we adopt a Gaussian noise in the number of estimated electrons with a deviation following a Poissonian distribution, that is $\varepsilon_{E,i}\propto\sqrt{E_i}$. The error in the number of electrons is propagated  through equation \ref{e:numbere} to the error in the flux as 
\begin{equation}
\varepsilon_{\phi,i}=\frac{1}{\lambda_i}\sqrt{\frac{\phi_i}{\mathcal{C}}}
\end{equation}
For a fixed SNR defined as the quotient of \ref{e:S} over \ref{e:N}, we can derived the normalization constant $\mathcal{C}$ as
\begin{equation}
    \mathcal{C}=({\rm SNR})^2\frac{\sum_i^n\phi_i/\lambda_i^2}{\sum_i^n\phi_i^2}.
\end{equation}
 
 Assuming that the SNR is proportional to the flux, SNR $\propto \phi$ and  using the relation between magnitude (in our case the {\it Gaia} magnitude $G$) and flux,

\begin{equation}
    G = -2.5 \cdot \log(\phi)+C,
\end{equation}
where $C$ is a constant of calibration, we can easily derive the relation between SNR and magnitude: 

\begin{equation}
\label{e:SNRG}
    {\rm SNR}_2 = {\rm SNR}_1 \cdot  10^{\frac{G_1-G_2}{2.5}}.
\end{equation}

From Table\,\ref{tab:gaia_SNR} the values of the SNR for $G = 13$ are known.  Thus, by adopting $G_1$=13 and the corresponding SNR$_1$ depending on the type of star and the range of wavelengths (BP or RP), we can estimate the SNR$_2$ for each one of the objects of magnitudes $G_2$. For  binary stars we only have approximations of SNR for their MS and WD components individually. In this case, we assume that individual fluxes are of the same order, $\phi_1\approx\phi_2$, which leads to signals of the same order, $S_1\approx S_2$ and results in a combined SNR of the system as
\begin{equation}
\label{e:SNR1SNR2}
{\rm SNR}=2\frac{{\rm SNR_1}\cdot{\rm SNR_2}}{{\rm SNR_1}+{\rm SNR_2}}
\end{equation}
where ${\rm SNR_1}$ and ${\rm SNR_1}$ are the individual SNR.


\end{document}